\documentclass[pdflatex,sn-mathphys-num]{sn-jnl}
\usepackage{graphicx}%
\usepackage{multirow}%
\usepackage{amsmath,amssymb,amsfonts}%
\usepackage{amsthm}%
\usepackage{mathrsfs}%
\usepackage[title]{appendix}%
\usepackage{xcolor}%
\usepackage{textcomp}%
\usepackage{manyfoot}%
\usepackage{booktabs}%
\usepackage{algorithm}%
\usepackage{algorithmicx}%
\usepackage{algpseudocode}%
\usepackage{listings}%
\usepackage{lmodern}
\usepackage{anyfontsize} 
\usepackage{enumerate} 
\usepackage{booktabs}
\usepackage{rotating}
\usepackage{makecell}
\begin{document}

\title[Beginner's Charm]{Beginner's Charm: Beginner-Heavy Teams Are Associated With High Scientific Disruption}

\author[1]{\fnm{Mahdee Mushfique} \sur{Kamal}}\email{mahdee.m.kamal@gmail.com}
\author*[2]{\fnm{Raiyan Abdul} \sur{Baten}}\email{rbaten@usf.edu}

\affil[1]{\orgdiv{Department of Computer Science and Engineering}, \orgname{Bangladesh University of Engineering and Technology}, \orgaddress{\city{Dhaka-1000}, \country{Bangladesh}}}

\affil[2]{\orgdiv{Bellini College of Artificial Intelligence, Cybersecurity, and Computing}, \orgname{University of South Florida}, \orgaddress{\street{4202 E Fowler Avenue}, \city{Tampa}, \state{FL-33620}, \country{USA}}}

\abstract{Teams now drive most scientific advances, yet the impact of absolute beginners---authors with no prior publications---remains understudied. Analyzing over 29 million articles published between 1941 and 2020 across disciplines and team sizes, we uncover a near-universal and previously undocumented pattern: teams with a higher fraction of beginners are systematically more disruptive and innovative. Their contributions are linked to distinct knowledge-integration behaviors, including drawing on broader and less canonical prior work and producing more atypical recombinations. Collaboration structure further shapes outcomes: disruption is high when beginners work with early-career colleagues or with co-authors who have disruptive track records. Although disruption and citations are negatively correlated overall, highly disruptive papers from beginner-heavy teams are highly cited. These findings reveal a ``beginner’s charm'' in science, highlighting the underrecognized yet powerful value of beginner fractions in teams and suggesting actionable strategies for fostering a thriving ecosystem of innovation in science and technology.}


\maketitle

\section{Introduction}\label{intro}
Imagine a research team with a high fraction of ``beginner'' authors---individuals with no prior publication. How likely is this team to disrupt its field? Intuitively, one might expect such teams to be unlikely engines of discovery. Scientific creativity rarely arises ex nihilo; theories of scientific and technological change emphasize discovery as an endogenous, cumulative process in which past knowledge enables future progress~\cite{fleming2001recombinant, schilling2011recombinant,fleming2010lone,baten2022novel,baten2021cues,baten2020creativity,kelty2025innovation,baten2024ai}. In this process, extended training and accumulated expertise allow researchers to stand, in Newton’s phrase, ``on the shoulders of giants''~\cite{acemoglu2016innovation, mukherjee2017nearly,kuhn1997structure}. Indeed, combinatorial accounts of creativity emphasize atypical recombinations of such well-established knowledge components as the substrate of breakthroughs~\cite{wang2017bias, lee2015creativity}. Moreover, as fields advance, the ``burden of knowledge'' intensifies, requiring ever longer training to reach the frontier and leaving less time to extend it~\cite{jones2009burden, jones2011age}. By this logic, beginner authors---who have had little time to digest this burden---may be handicapped in their capacity for field-shifting innovation and disruption. If mastery is required before divergence, then beginner-heavy teams may lack the cognitive materials and experiential grounding to engineer novelty.

Yet, there are equally compelling reasons to expect that beginner-heavy teams may, in fact, be well-positioned to disrupt scientific advancement. Youth is historically linked to novelty in science: they are less encumbered by entrenched paradigms, more open to new ideas, and more willing to take intellectual risks~\cite{rappa1993youth, packalen2019age}. Freed from institutionalized pressures---such as administrative burdens and funding constraints that often steer senior researchers toward conservative publication strategies~\cite{cohen1985revolution, spreng2021exploration, livan2019don, fortunato2018science,simonton2014wiley,higashide2024mid, spencer2017research, mcalpine2018identity, schimanski2018evaluation}---beginners may pursue bold ideas more readily. Moreover, deep disciplinary expertise can harden into attachment, making it harder for experienced scientists to selectively ``unlearn'' prevailing assumptions and embrace radically new ones~\cite{cui2022aging,merton1973sociology, candia2021quantifying, lin2025disruption}. From this perspective, beginners' absence of enculturation may be a strength precisely because they have not yet fully assimilated the burden of knowledge. Indeed, many landmark discoveries originated as debut publications---such as Turing Awardee Yann LeCun’s early doctoral work, which laid the foundation for backpropagation~\cite{lecun85, lecun_publis_2025}. Extending this logic to the team level, beginner-heavy groups may be especially disruptive: their fresh cognitive repertoires, freedom from orthodoxy, and high risk tolerance may outweigh individual knowledge deficits, enabling breakthrough contributions more readily than in teams with fewer beginners.

While recent studies have explored the aggregated effects of mean career age on team disruption~\cite{yang2024unveiling,zeng2021fresh}, the role of the very youngest contributors---beginner authors---remains overlooked. In particular, it remains unclear how the disruptivity of teams changes as the fraction of beginners increases, and what underlying team dynamics and knowledge integration mechanisms drive these outcomes. To address this gap, we conduct a large-scale study of more than $29$M articles published between $1941$ and $2020$ across $19$ broad scientific disciplines. Our analysis draws on SciSciNet V2~\cite{lin2023sciscinet, sciscinet-v2}, one of the most comprehensive and widely used bibliometric data lakes. Using SciSciNet's name-disambiguated author data~\cite{torvik2009author}, we calculate career age as the number of years since an author’s first publication~\cite{frandsen2024defining, bradshaw2021fairer,cui2022aging}, and classify authors into three career stages for our main results: beginner (0 years), early-career (1–10 years), and senior researchers (11 or more years). These thresholds reflect typical academic transitions, with the early stages encompassing graduate training and initial faculty appointments, and seniority being attained after a decade of publishing~\cite{vilela2023career, bonaccorsi2003age}. We also conduct robustness tests using a different choice of career stage thresholds: beginner (0 years), early-career (1–5 years), mid-career (6-10 years), and senior researchers (11 or more years) to confirm the generality of the results.

We find strong evidence for a near-universal and previously undocumented pattern of ``beginner's charm'': an increased presence of beginner authors in teams is associated with higher scientific disruption and innovation. This association holds across disciplines, time periods, and team sizes, suggesting it reflects a general feature of how science advances rather than a peculiarity of specific fields or historical moments. 

\begin{figure*}[t]
\centering
\includegraphics[width=1\linewidth]{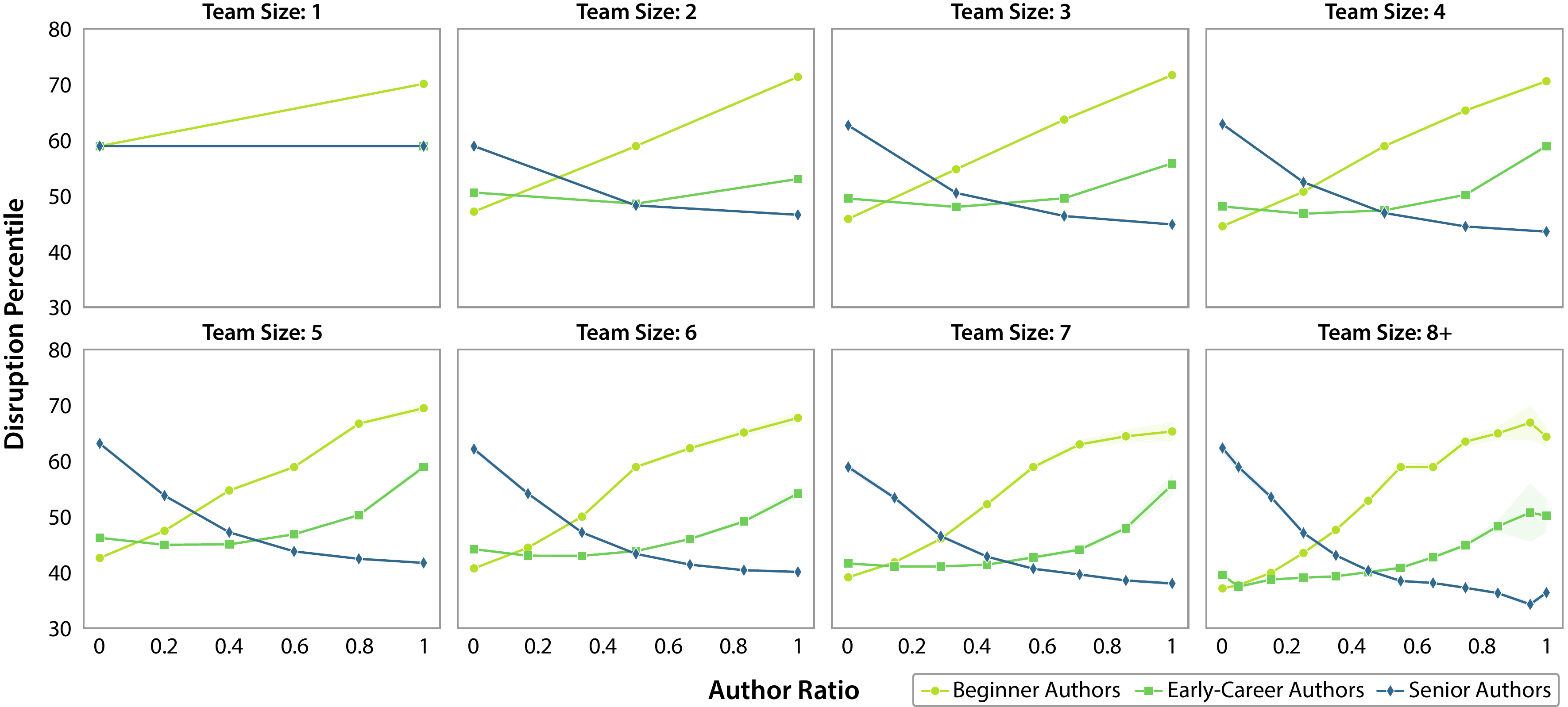}
\caption{\textbf{Beginner-heavy teams are robustly disruptive across team sizes.} We show the relationship between disruption percentiles and author ratios at different career stages, split by team sizes. Beginner author ratios are positively associated with higher levels of disruption, while teams with higher senior author ratios tend to produce less disruptive work. The results are robust to team size. Shaded regions denote $95$\% C.I.}
\label{career_age_disruption}
\end{figure*}

\section{Results}\label{results}
\subsection{Teams with higher beginner-author ratios are more disruptive and innovative} 
As our primary team performance metric, we use the disruption score, originally developed to distinguish between consolidating and destabilizing contributions in science and technology~\cite{funk2017dynamic}. This score ranges from $-1$ to $+1$, capturing papers that reinforce or shift scientific trajectories, respectively (see Materials and Methods). Since its introduction, the disruption score has been widely adopted and validated across domains~\cite{wu2019large, lin2023sciscinet, leibel2024we, chu2021slowed, park2023papers, wei2023quantifying}, providing a robust lens for answering our research queries.

We find that papers with a higher proportion of beginner authors tend to be more disruptive: in the overall dataset, the beginner ratio is positively correlated with a paper's disruption percentile (Pearson’s $r=0.11$, Holm-adjusted $P<10^{-6}$), while early-career ratio shows a small positive association ($r=0.002$, $P<10^{-6}$), and senior-author ratio shows a negative association ($r=-0.08$, $P<10^{-6}$).

While the overall correlations are informative, they obscure the role of team size, since beginner ratios are mechanically constrained by the number of authors and disruption itself is known to vary with team size~\cite{wu2019large}. To address this, we stratify the analysis by team size and recompute the associations (Figure~\ref{career_age_disruption}). The results are strikingly monotonic across team sizes: beginner ratios show uniformly strong positive correlations with disruption (Kendall’s $\tau$ up to 1.0 with narrow 95\% CIs, $P<0.001$ in larger teams), senior ratios show consistently negative associations (often $\tau=-1.0$, $P<0.001$), and early-career ratios display weaker but positive effects that reach significance in larger teams (Supplementary Table~\ref{SItab_s1}; all $P$-values are Benjamini–Hochberg FDR-corrected).

Prior studies found that larger teams tend to produce less disruptive work~\cite{wu2019large}, and we replicate this pattern in teams with low beginner ratios. Interestingly, however, the positive effect of beginner ratio on disruption is especially pronounced in larger teams, where the usual size penalty attenuates (Supplementary Figure~\ref{SIfig_s1}). As a result, beginner-heavy large teams approach the disruption levels of beginner-heavy small teams. One possible explanation is that a higher beginner ratio may counteract the convergence pressures of large teams—such as reduced risk-taking~\cite{christensen2011innovators}, resistance to external perspectives~\cite{minson2012cost}, and neutralized viewpoints~\cite{greenstein2016open}—enabling a shift from consolidating “safe bets” to more exploratory, high-variance work~\cite{paulus2013understanding,lakhani2013prize}.

The positive association between beginner share and disruption is robust across all eight decades and in all 19 broad academic disciplines in the corpus ($r > 0$, $P < 0.001$ throughout; Supplementary Tables~\ref{SItab_s2}-\ref{SItab_s3}). Decade-specific analyses show the positive association to hold true across team sizes ($r>0$, $P<10^{-6}$ throughout; Supplementary Figure~\ref{SIfig_s2}). For discipline-specific analysis, we leverage SciSciNet’s field-of-study annotations derived from the Microsoft Academic Graph (MAG), which organizes research into a hierarchical taxonomy of fields~\cite{lin2023sciscinet,sinha2015overview}. In particular, we use SciSciNet's 19 Level-0 (top-level) field annotations as broad disciplines, allowing each paper to be associated with one or more disciplines. We include all papers associated with a given discipline, such that each multidisciplinary paper contributes to multiple disciplines. Across all 19 disciplines, the positive association between beginner share and disruption remains robust across team sizes (Supplementary Figure~\ref{SIfig_s3}). Collectively, these patterns indicate that disruption is robustly elevated in teams with a higher fraction of beginners.

To verify that our findings are not artifacts of the disruption score metric, we employ a secondary metric of atypical combination scores (see Materials and Methods)~\cite{jones2009burden,uzzi2013atypical,guimera2005team,jones2008multi}. This score conceptualizes discovery as innovative knowledge-recombination: for each paper, all pairs of reference venues are scored for atypicality against a degree- and time-preserving null, and the median $z$-score of all these pairs is taken as the paper’s atypical combination score. A score below $0$ represents innovative combinations of existing knowledge, and vice versa~\cite{uzzi2013atypical}. Unlike disruption (a downstream impact measure), the atypical combination score captures innovation from a paper's \emph{inputs}. 

\begin{figure}[t]
\centering
\includegraphics[width=0.65\linewidth]{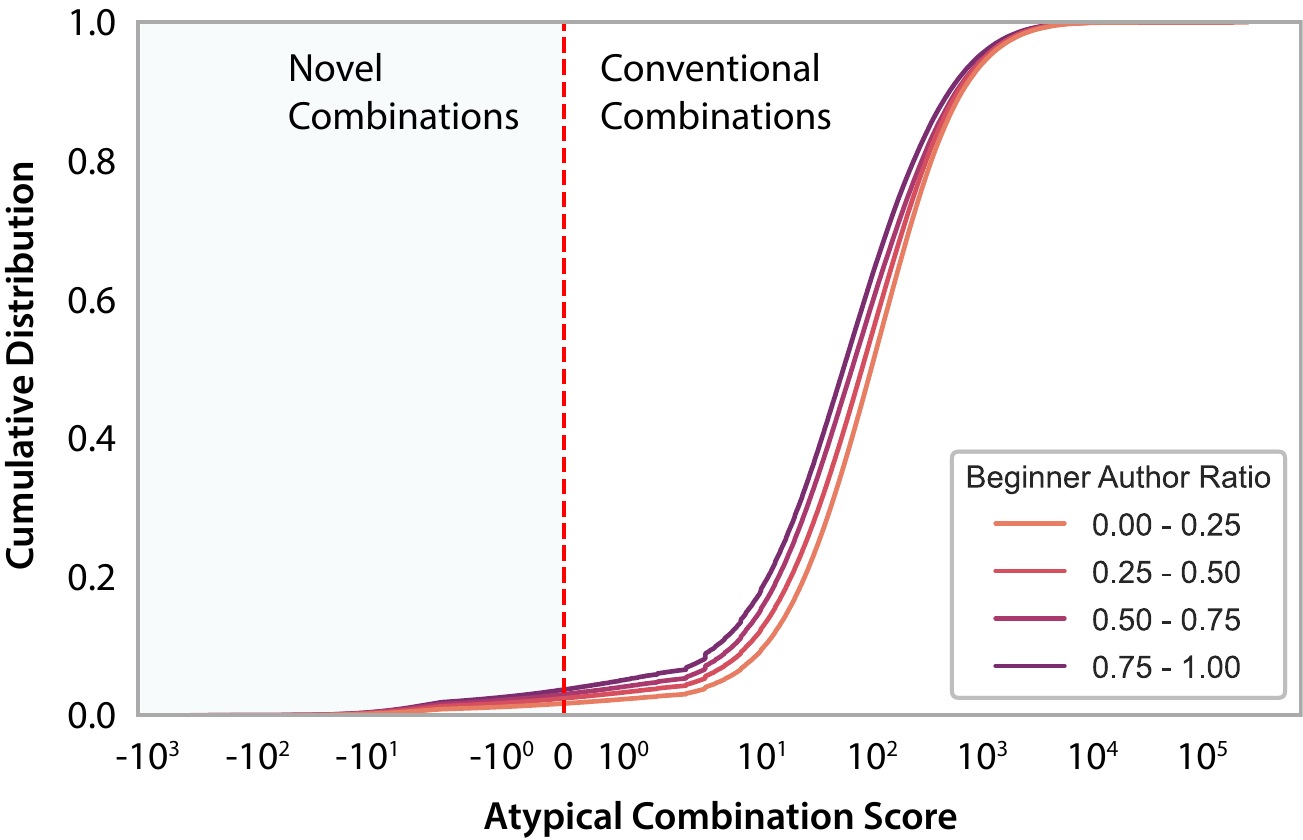}
\caption{\textbf{Beginner-heavy teams are innovative.} We show the cumulative distributions of atypical combination scores across beginner-author ratio quartiles. Negative atypical combination scores indicate innovative combinations, and positive scores indicate conventional ones. At higher beginner author ratio quartiles, innovative combinations are systematically more prevalent.}
\label{atyp}
\end{figure}

Figure~\ref{atyp} shows the cumulative distributions of atypical combination scores for different beginner author-ratio quartiles (Q1: 0–0.25, Q2: 0.25–0.50, Q3: 0.50–0.75, Q4: 0.75–1.00). As the share of beginner authors increases, the cumulative distributions shift toward lower values, indicating more innovative combinations. The ECDFs of the atypical combination scores differ significantly between every quartile pair (two-sample Kolmogorov–Smirnov test; Holm-adjusted $P<10^{-300}$ for all comparisons; largest contrast Q1 vs Q4: $D=0.142$). Supplementary Figure~\ref{SIfig_s4} shows that higher proportions of beginner authors are associated with lower atypicality score percentiles (higher innovation). In contrast, higher senior-author ratios are associated with higher atypicality score percentiles (lower innovation). Together, these findings corroborate and validate our disruption-based results that beginner-heavy teams are more innovative than beginner-light ones.

The marked difference in disruption and innovation between beginner-heavy and other teams prompts a closer examination of how beginner-heavy teams build on prior work. Our atypical combination score-based results reveal one nuance: such teams draw from an unconventionally broader range of prior knowledge domains, as reflected in their use of unusual reference-venue pairings. To further probe this, we examine two metrics: reference popularity (mean citations of a paper’s references; higher values reflect citing ``hotter'' work) and reference age (mean year gap between a paper and its references; lower values reflect more recent sources). 

We find that the proportion of beginner authors is negatively associated with reference popularity (Kendall’s $\tau$ up to $-1$ for all team sizes, Benjamini–Hochberg FDR-corrected $P < 0.05$ for team sizes $\ge 5$; Supplementary Figure~\ref{SIfig_s5}, Supplementary Table~\ref{SItab_s4}), but shows no systematic association with reference age. In other words, beginners gravitate toward less-cited work of similar vintage---potentially underexplored or neglected areas---consistent with a broader, less canonical knowledge base that can support novel recombinations. This pattern suggests that breadth may serve collective scientific progress better than chasing the highly cited canon, and beginner-heavy teams tend to tap into prior art more broadly.

Another key difference between beginners and later-career authors is productivity. Beginners publish fewer papers in their debut year---averaging $1.18$ in year $0$ versus $1.68$ for career-age $1$ authors---a significant increase of $+0.50$ (Kruskal–Wallis test, $P<0.001$) that persists across subsequent years, with only modest annual increases afterward (maximum $+0.13$; Supplementary Figure~\ref{SIfig_s6}). This structural difference suggests that beginners initially emphasize novelty over volume, echoing prior work showing that higher productivity can reduce disruption~\cite{li2024productive,park2023papers,chu2021slowed,li2022bibliographic,michalska2017and,li2025quantifying}. Taken together, our results suggest that beginners tend to favor breadth over quantity in ways that are beneficial for disruption but may become challenging to sustain later in their careers.

\begin{figure*}[t]
\centering
\includegraphics[width=1\linewidth]{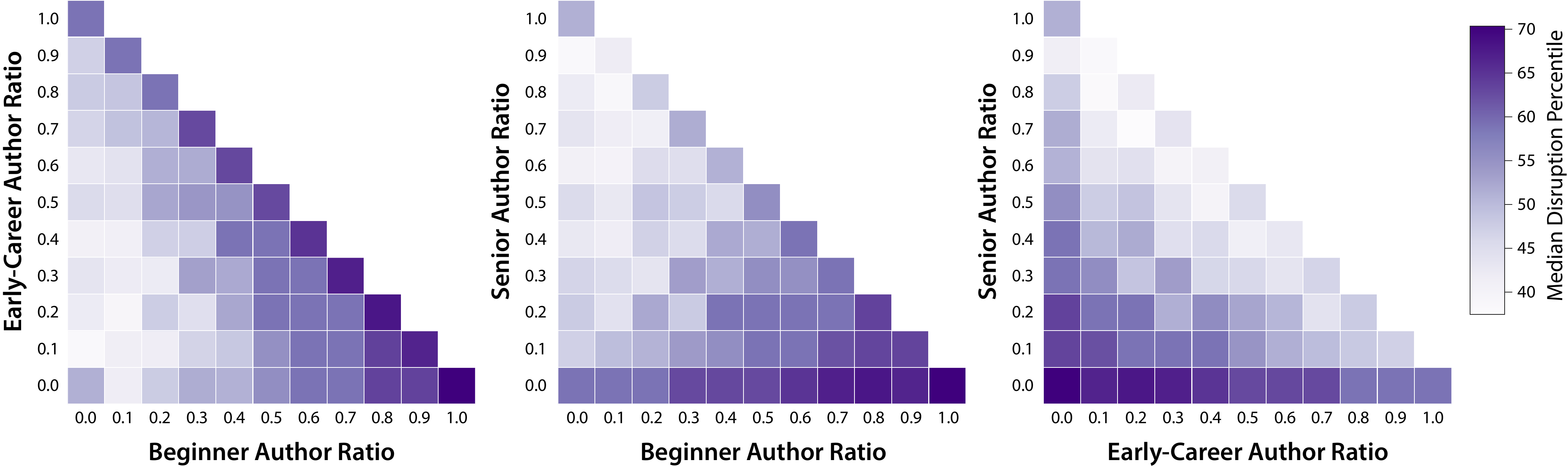}
\caption{\textbf{Early-career collaborators are associated with higher disruption in beginner-heavy teams.} We illustrate disruptions of pairwise combinations of beginner, early-career, and senior authors across different author ratios. The heatmap cells visualize median disruption percentiles, with darker shades indicating higher disruption. While teams heavier on beginners are progressively more disruptive, early-career collaborators are associated with better beginner disruption than senior collaborators. See Supplementary Tables~\ref{SItab_s5}-\ref{SItab_s7} for more details.}
\label{disruption_heatmap}
\end{figure*}

\begin{figure}[t]
\centering
\includegraphics[width=0.85\linewidth]{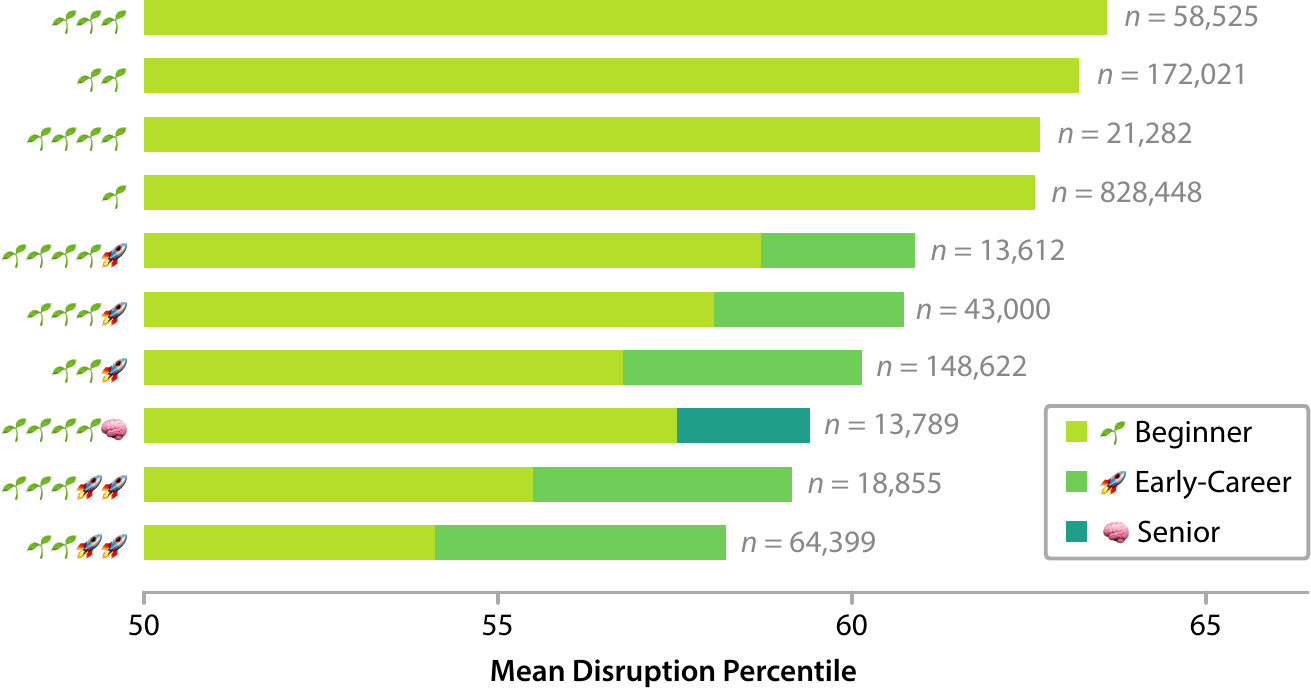}
\caption{\textbf{The most disruptive team compositions are beginner-heavy.} We show the top $10$ most disruptive team compositions in our dataset (ranked by mean disruption percentiles; minimum $n\ge10,000$ per composition). Each bar represents a unique team composition, with emojis indicating the exact number of authors from each career stage and colors denoting their proportional representation in the teams. Beginner authors systematically dominate the most disruptive compositions.}
\label{leaf}
\end{figure}

\subsection{Early-career and disruptive collaborators are linked to greater disruption in beginner-heavy teams}
How do non-beginner collaborators influence the disruptiveness of beginner-heavy teams? We examine disruption percentiles across pairwise combinations of beginner, early-career, and senior authors. For both early-career and senior collaborator settings, increasing the proportion of beginners is associated with higher disruption---peaking when beginners make up the entire team (Figure~\ref{disruption_heatmap}, Supplementary Tables~\ref{SItab_s5}-\ref{SItab_s7}). Notably, teams comprising beginners and early-career researchers show greater disruption than other compositions, suggesting that younger teams are especially conducive to disruptive breakthroughs. These patterns are replicated using atypicality scores (Supplementary Figure~\ref{SIfig_s7}).

Next, we examine prior disruption records of the non-beginner collaborators. We compute the average disruption scores of the prior publications of each early-career and senior collaborator. Using this, we rank teams based on the average disruptiveness of the non-beginner collaborators. We find that teams that bring together beginner authors and more disruptive co-authors tend to produce more disruptive work. As shown in Supplementary Figure~\ref{SIfig_s8}, teams with an equal fraction of beginner authors become more disruptive as the collaborators' average disruption rank increases, robustly across different beginner author ratios and team sizes. Together, these results highlight the importance of younger and highly disruptive collaborators in shaping the beginners' contributions.

Finally, we examine beginner ratios in the highly prevalent and highly disruptive team compositions in our dataset. We isolate high-frequency team compositions (each with $\ge 10,000$ papers and team size $\le 40$) to identify $103$ unique compositions of beginner, early-career, and senior author counts. We rank these compositions by their mean disruption percentiles (Figure~\ref{leaf}). Remarkably, beginner-rich teams remain prominent: among the top $50$ most disruptive compositions, $24$ have at least as many beginners as other career stages, and in $15$ cases, beginners outnumber all other career stages combined. Crucially, this effect is not a fluke: each composition reflects \textit{at least} $10,000$ papers over eight decades, revealing a robust, recurring beginner advantage with consistently high disruption.

\begin{figure}[t]
\centering
\includegraphics[width=0.65\linewidth]{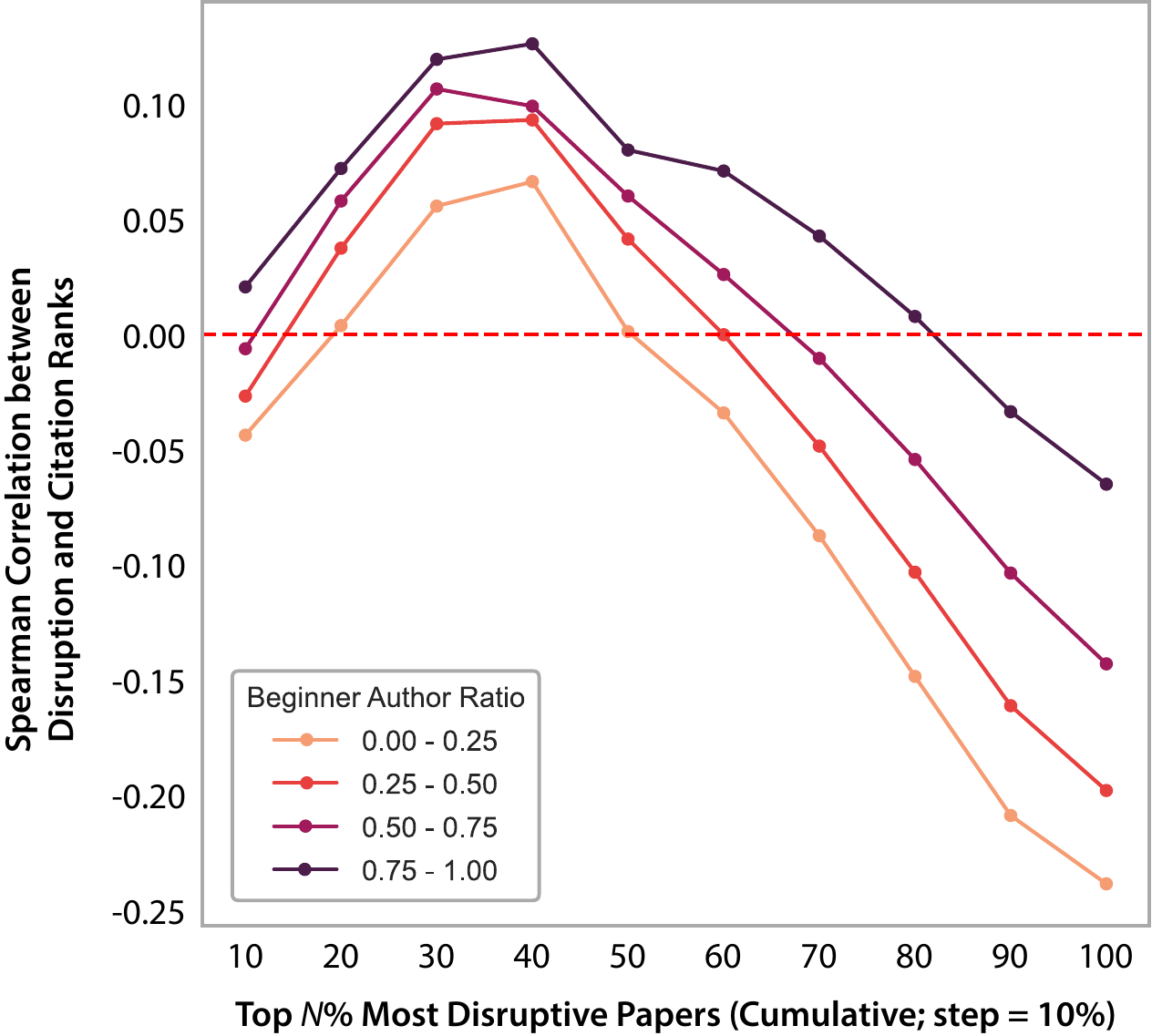}
\caption{\textbf{Highly disruptive papers by beginner-heavy teams are highly cited.} We show the correlation coefficients between disruption and citation ranks across various cumulative deciles (top $N$\%) of disruptive papers, split by beginner-ratio quartiles. A positive correlation implies that disruptiveness is rewarded with citation impact, and vice versa. As teams grow beginner-heavy, their trajectories of correlation coefficients remain positive for a larger portion of the top-disruption deciles.}
\label{citation_corr}
\end{figure}

\subsection{Highly disruptive papers by beginner-heavy teams are highly cited}
How well is disruptive work by beginner-heavy teams received by the academic community? Citations are commonly used as a proxy for academic impact~\cite{sinatra2016quantifying}. Prior work documented a negative overall relationship between disruption scores and citation counts~\cite{li2024breaking,zeng2023disruptive}, a result corroborated in our dataset. Unsurprisingly, we observe an overall negative association between the proportion of beginner authors in teams and their citation impact (Supplementary Figure~\ref{SIfig_s9}).

However, this aggregate trend obscures a more nuanced citation dynamic among the most disruptive papers~\cite{li2024breaking}. To uncover it, we first rank all papers by disruption scores and citation counts. Then, for each cumulative top-$N$\% of disruptive papers ($N=10$ to $100$, in steps of $10$), we compute the Spearman correlation between disruption and citation ranks within each beginner-ratio quartile (Q1–Q4). A positive correlation implies that high disruptiveness is rewarded with high citation impact, and vice versa.

Figure~\ref{citation_corr} reveals that among the most disruptive papers (smaller $N$), only the highest beginner-ratio quartile (Q4) consistently shows a positive disruption-citation correlation. As $N$ increases, correlations across all quartiles rise to the positive territory, peaking around the $N=40\%$ mark. Beyond that, correlations decline and eventually become negative for all quartiles by $N=100\%$, mirroring the negative trend observed in the full dataset. In short, the most disruptive papers from beginner-heavy teams tend to be highly cited; however, this advantage weakens as less disruptive papers are included in the cumulative subsets. To explain these results further, Supplementary Figure~\ref{SIfig_s10} illustrates the citation fractions received by non-cumulative top-$N$\% bins of disruption-ranked works. Highly disruptive papers (smaller $N$) receive a disproportionately high share of citations (i.e., higher than the random baseline of all deciles receiving $10\%$ share of all citations). In contrast, mid-range disruptive papers (especially around $N=40\%$) are the least cited. The citation share then increases again for more consolidating work with lower disruption scores. This non-monotonic distribution helps explain why the correlations turn negative at scale.

Importantly, teams with more beginners (Q4) maintain positive disruption–citation correlations for the broadest top-disruption range, whereas beginner-sparse teams (Q1) exhibit the fastest drop-off into negative territory (Figure~\ref{citation_corr}). These quartile differences are robust across all cumulative top-$N$ thresholds: pairwise Fisher $r$-to-$z$ tests (Holm-adjusted) reveal significant divergence between every quartile pair at every threshold ($P<10^{-300}$). The most pronounced gap appears between Q1 and Q4, maximizing at $N=100\%$ with a Fisher-transformed $z$-score difference of $\Delta z = -185.91$ ($P<10^{-300}$). Together, these findings suggest that while disruptive work is generally harder to recognize, highly disruptive papers from beginner-heavy teams are well-rewarded with high citation impact. This supports the notion that newcomers can succeed by taking bold intellectual leaps.

\subsection{Robustness to alternative career-stage definitions}
As a robustness check, we refine our career-stage definitions by splitting early-career authors into early (1–5 years) and mid-career (6–10 years) groups, corresponding roughly to graduate training versus early faculty stages, while keeping the beginner and senior definitions unchanged. The beginner category (0 years) is fixed by construction, as it captures authors with no prior publications. The senior threshold (11+ years) reflects a decade or more of publishing, a period reasonably sufficient for disciplinary enculturation and the accumulation of domain-specific expertise. We therefore focus the refinement on the intermediate career range, where heterogeneity in training stage and professional autonomy is most pronounced. 

This alternative specification yields the same qualitative patterns across the main results. In the overall dataset, the beginner ratio remains positively associated with disruption (Pearson’s $r=0.11$), the early-career ratio is much weaker ($r=0.005$), the mid-career ratio is slightly negative ($r=-0.002$), and the senior ratio remains most negative ($r=-0.08$), preserving the same graded decline across career stages from beginners to seniors. The team-size-stratified analyses likewise preserve the ordinal pattern: beginner ratios remain most positively associated with disruption across team sizes, early- and mid-career ratios show weaker intermediate associations, and senior ratios remain negative (Supplementary Figure~\ref{SIfig_s11} and Supplementary Table~\ref{SItab_s8}).

The collaborator-composition analyses are similarly stable under the four-way partition: the disruption heatmaps continue to show that teams become more disruptive as the beginner ratio rises, while the corresponding atypical-combination heatmaps preserve the same directional ordering of innovation-related patterns across collaborator mixes (Supplementary Figures~\ref{SIfig_s12}--\ref{SIfig_s13}). Taken together, these checks show that our conclusions do not depend on treating years 1--10 as a single class, but instead reflect a stable career-stage gradient in which beginner-heavy teams remain the most disruptive and innovative.

\section{Discussion}
Teams are today the dominant engine of science, spanning nearly every field of inquiry and steadily replacing the lone researcher in driving knowledge forward~\cite{wuchty2007increasing, jones2008multi}. A central challenge in the science of science is to uncover what kinds of team dynamics foster originality and breakthroughs. Here we provide evidence for a near-universal and previously undocumented pattern: the presence of beginners---authors with no prior publication history---systematically enhances scientific disruption and innovation. This finding adds a critical new dimension to our understanding of how knowledge advances, challenging the prevailing assumption that beginners are merely blank slates or passive contributors. 

Our insights were made possible by a large-scale analysis of $29$ million papers across eight decades. We find that beginner-heavy teams consistently produce more disruptive and innovative work across disciplines, time periods, team sizes, and alternative definitions of career stages. This success is linked to distinctive knowledge-integration behaviors: beginners are less tethered to prevailing trends, more likely to pursue neglected or atypical references, and thus better positioned to recombine knowledge in novel ways. These mechanisms align with longstanding theories of creativity that emphasize originality through unusual combinations and broader search~\cite{uzzi2005collaboration, schilling2005small, jones2009burden,schoenmakers2010technological, kelley2013breakthroughs, didegah2013factors, mammola2021impact}. While philosophers of science describe knowledge growth as an endogenous process fueled by accumulated understanding, it requires engagement with a wide range of prior work—a condition beginners appear to fulfill despite having less time to absorb the field’s expanding burden of knowledge~\cite{jones2009burden}. Unburdened by entrenched norms, beginners may be more open to exploring conceptual recombinations that others overlook. Their lower productivity relative to later-career researchers may also allow for more focused intellectual risk-taking, aligning with creativity theories that emphasize deep attention and cognitive agility as drivers of novel contributions~\cite{li2024productive,park2023papers,chu2021slowed}.

Collaboration dynamics also play a crucial role in shaping beginner success. We find that early-career collaborators are associated with greater disruption in beginner-heavy teams than senior ones, suggesting that younger collaborators may offer more compatible forms of intellectual flexibility and support. We additionally find that beginner-heavy teams thrive in disruptiveness when paired with collaborators who themselves have disruptive track records. These results contribute to a growing literature emphasizing how team composition, career age, diversity, and freshness~\cite{wu2019large, zeng2021fresh, hofstra2020diversity, yang2024unveiling,packalen2019age, kwiek2024young,grant1996toward, kunze2011age, schneid2016age, li2021leveraging} mediate collaborative innovation. Encouragingly, we find that the highly disruptive works by beginner-heavy teams are rewarded with high citation impact in the academic literature, corroborating and further elucidating prior findings~\cite{li2024breaking}.

Our work provides deeper nuances to contextualize previously reported insights on team disruption. For instance, prior work has shown that larger teams tend to be less disruptive than small ones~\cite{wu2019large}, that team performance follows a U-shaped relationship with average career age~\cite{yang2024unveiling}, and that ``fresh'' teams (i.e., those without prior collaborations among members) are more likely to produce original or multidisciplinary work~\cite{zeng2021fresh}, among others. However, the role of absolute beginners with no prior publications has remained largely unexamined. To that end, we identify a pathway through which the well-known penalty of large team size on disruptiveness can be negated: \textit{beginner-heavy} large teams can rival their smaller counterparts in disruptive capacity. Moreover, our results offer a finer-grained understanding of the left arm of the U-shaped curve relating average career age and disruption of teams~\cite{yang2024unveiling}, revealing the previously undocumented importance of the presence of absolute beginners in the low average career age territory. This is especially relevant given that over $98.2\%$ of scientific teams fall on the left side of that U-curve, underscoring the widespread applicability of our findings. We also extend earlier findings about team freshness by showing that not only do fresh \textit{collaborations} enhance disruption by bringing in diverse expertise to the team, but so too does \textit{individual-level} freshness, who bring in no track record. This complements and deepens prior understandings of team novelty~\cite{zeng2021fresh}, offering a more granular account of how new entrants inject creative potential into scientific teams.

Our study has limitations. We define beginners solely by career age, without distinguishing differences in physical age or the amount of prior training. We also adopt a team-level lens, without tracking individual beginners longitudinally to evaluate whether their disruptive success persists or peaks later in their careers~\cite{liu2018hot, li2025quantifying,wang2019early,yin2019quantifying,liu2021understanding}. Future work should explore these trajectories, as well as the micro-dynamics of collaboration: how beginner ideas are introduced, evaluated, and either nurtured or sidelined. Moreover, examining whether similar patterns hold in domains such as startups, patents, or the arts could reveal whether the ``beginner's charm'' generalizes beyond academia. While the disruption and atypical combination scores we used capture fundamentally different dynamics---paradigm-shifting effects in \textit{downstream} literature versus novelty in recombining \textit{prior} literature, respectively~\cite{wang2023effect,leibel2024we,leahey2023types,funk2017dynamic}---future work can apply additional indicators of innovation, including natural language processing-based measures to analyze abstract-level novelty, conceptual distance, or idea emergence over time. Importantly, our findings are limited to correlational observations rather than making causal claims.

The implications of these findings are wide-ranging and challenge the prevailing emphasis on experience and track record in science hiring, funding, and team assembly. \textit{First}, they offer guidance for team science: teams designed for disruptive outcomes may benefit from intentionally integrating beginners alongside collaborators with disruptive track records. This stands in tension with two ongoing trends. On one hand, industry hiring is increasingly privileging elite track records—exemplified by Meta’s recent push to “build the most elite and talent-dense team in the industry”~\cite{duffy2025meta}. On the other, AI is disproportionately displacing entry-level roles, reducing traditional pathways through which newcomers gain experience and contribute intellectually~\cite{wef2026_entry_level_ai}. Together, these forces risk systematically excluding precisely the beginner contributions that our results identify as critical for innovation. \textit{Second}, at the institutional level, universities and research organizations might reconsider evaluation systems overly tied to citation counts and productivity (i.e., the so-called ‘publish or perish’ culture~\cite{plume2014publish}), shifting toward measures that reward originality, risk-taking, and long-term vision~\cite{reisz2022loss}. \textit{Finally}, at the policy level, funding agencies could create programs that explicitly support beginner-led projects or incentivize collaborations that amplify beginner contributions. Such efforts would counteract systemic biases that currently privilege safe, incremental research over risky exploration~\cite{rzhetsky2015choosing, livan2019don}.

Overall, our results highlight a form of “beginner’s charm” in science: newcomer-heavy teams occupy a unique position to disrupt, driven by their fresh perspectives and willingness to explore less-traveled intellectual paths. As science becomes more collaborative, and as the academic workforce ages~\cite{milojevic2018changing,jones2009burden,matthews2011aging,sugimoto2016age,blau2017us}, the under-tapped value of newcomers becomes increasingly vital to shape the future trajectory of discovery.

\section{Materials and Methods}\label{methods}
\subsection{Dataset}
In this paper, we analyze SciSciNet V2, a refreshed update to SciSciNet. It is a large-scale, integrated data lake derived from the Microsoft Academic Graph (MAG)~\cite{wang2020microsoft} and OpenAlex \cite{openalex} to support science of science research. The dataset contains information on 249,803,279 publication records. The author names are disambiguated in the database~\cite{torvik2009author}. We select all articles published between 1941 and 2020 across all 19 Level-0 (top-level) disciplines. Furthermore, we remove extreme entries (authors with $1000$ or more overall publications, authors whose career span is $80$ or more, authors who publish more than $50$ papers in a year, and beginner authors with $10$ or more papers in their debut year) to curate a final dataset of 29,054,261 papers for this analysis. However, all of our findings remain true even if we include these extreme entries. Please note that we do not include earlier decades due to a lack of samples available (e.g., after filtration, only 2037 papers remain between 1931-1940, and even fewer for earlier decades). We explain the dataset preprocessing steps in the Supplementary Text.

\subsection{Disruption Score} 
Citation networks are a standard tool for measuring scientific disruption---identifying works that introduce new directions rather than build incrementally on prior research. The Disruption Score $D$ quantifies this by comparing how future papers cite the focal paper and its references~\cite{funk2017dynamic, wu2019large}:
\begin{equation*}
D = \frac{N_i - N_j}{N_i + N_j + N_k}.
\end{equation*}
Here, $N_i$ is the number of papers that cite the focal paper but not its references, $N_j$ cite both, and $N_k$ cite only the references. The score ranges from $-1$ (developmental) to $+1$ (disruptive). Positive values suggest the focal paper has shifted the research trajectory, while negative values indicate work that deepens existing lines. We use the Disruption Score provided in the SciSciNet dataset~\cite{lin2023sciscinet}.

\subsection{Atypical Combination Score}
Building on the view that scientific advances arise from recombining existing knowledge~\cite{jones2009burden,uzzi2013atypical,guimera2005team,jones2008multi,yang2022gender}, we quantify atypicality following Uzzi et al.~\cite{uzzi2013atypical}. For each focal paper, all unordered pairs of reference venues in its reference list are formed, and the typicality of each pair is assessed using the literature preceding the paper’s publication year. Let $O_{ij}$ denote the observed prior co‐occurrence count of venues $i$ and $j$. $O_{ij}$ is compared to an empirical null generated by Monte Carlo edge‐switching on the global citation network that preserves (i)~each paper’s in‐ and out‐degree and (ii)~the forward/backward temporal distribution of citations, thereby holding constant the propensity of papers to cite and be cited over time. For each pair $(i,j)$, a null mean $\mu_{ij}$ and standard deviation $\sigma_{ij}$ are obtained across randomized networks, and a normalized score is computed:
\begin{equation*}
z_{ij} = \frac{ O_{ij} - \mu_{ij} }{ \sigma_{ij} }.
\end{equation*}
Here, positive $z_{ij}$ values indicate \emph{conventional} venue pairings (more frequent than expected by chance), whereas negative $z_{ij}$ values indicate \emph{atypical} venue pairings (less frequent than expected). A paper’s overall atypicality is summarized by the median of its pairwise $z$–scores (median $z$)~\cite{uzzi2013atypical}. Lower median $z$ values correspond to more atypical (novel) combinations; higher values indicate more conventional combinations. In this paper, we use the median $z$ values for papers provided by the SciSciNet data lake~\cite{lin2023sciscinet}, which computes $z$–scores using the procedure above with the prior‐year literature as the reference set. We restrict analyses to papers with at least two cited venues (so that at least one pair exists). We report the median $z$ as the ``Atypical Combination Score," with lower scores indicating greater novelty.

\backmatter
\bmhead{Acknowledgements}
A faculty startup fund from the University of South Florida supported this work.

\bmhead{Conflicts of Interest} The authors declare no conflict of interest.

\bmhead{Ethics Approval} This work uses public data and does not require IRB approval.

\bmhead{Data and Code Availability} The data is publicly available at \url{https://northwestern-cssi.github.io/sciscinet/}. Our end-to-end preprocessing and statistical analysis code is available at \url{https://github.com/cssai-research/beginners_charm}.

\bmhead{Author Contributions} MMK designed the study, analyzed the data, interpreted the results, and authored the manuscript. RAB designed the study, analyzed the data, interpreted the results, authored the manuscript, and supervised the research.

\clearpage

\begin{appendices}

\setcounter{figure}{0}
\setcounter{table}{0}
\renewcommand{\thefigure}{S\arabic{figure}}
\renewcommand{\thetable}{S\arabic{table}}

\makeatletter
\renewcommand{\theHfigure}{S\arabic{figure}}
\renewcommand{\theHtable}{S\arabic{table}}
\makeatother

\section*{Supplementary Information for \textit{Beginner's Charm: Beginner-Heavy Teams Are Associated With High Scientific Disruption}}\label{SM}

\subsection*{Data Preprocessing} We utilized BigQuery~\cite{google2025bigquery}, a Google Cloud Platform enterprise-level data warehouse tool, to preprocess SciSciNet data for our research. We first added a \texttt{debut\_year} column in the \texttt{Authors} table based on each author's earliest publication. After that, we constructed the \texttt{All\_Yearly\_Author\_Profiles} table, where \texttt{(authorid, year)} is a composite key. For a given year $y$, this table contains the \texttt{career\_age} of \texttt{authorid} in year $y$, the \texttt{number\_of\_papers} by \texttt{authorid} in year $y$, and aggregated metrics (productivity, average citation, average disruption, etc.) regarding their previous year's publications. Then, we enhanced the \texttt{Paper} table by adding pre-publication factors needed for our research. For each year $y$ ranging from $1941$–$2020$; and for each \texttt{paperid} published in year $y$, we looked up the rows for \texttt{(authorid, $y$)} to construct team-level pre-publication factors of a paper such as \texttt{beginner\_author\_count}, \texttt{beginner\_author\_ratio}, \texttt{average\_disruption\_of\_team}, and \texttt{average\_citation\_count\_of\_team}, by aggregating the retrieved rows for the \texttt{(authorid, $y$)} key for each author of the \texttt{paperid}. We discarded records where \texttt{doi} or \texttt{disruption} was missing.
Furthermore, we removed some unrealistic entries (described in the main manuscript). Finally, we used the rank function of Pandas~\cite{mckinney2010pandas} to calculate the disruption and citation percentiles of each paper.

\clearpage

\begin{figure}
\centering
\includegraphics[width=0.7\linewidth]{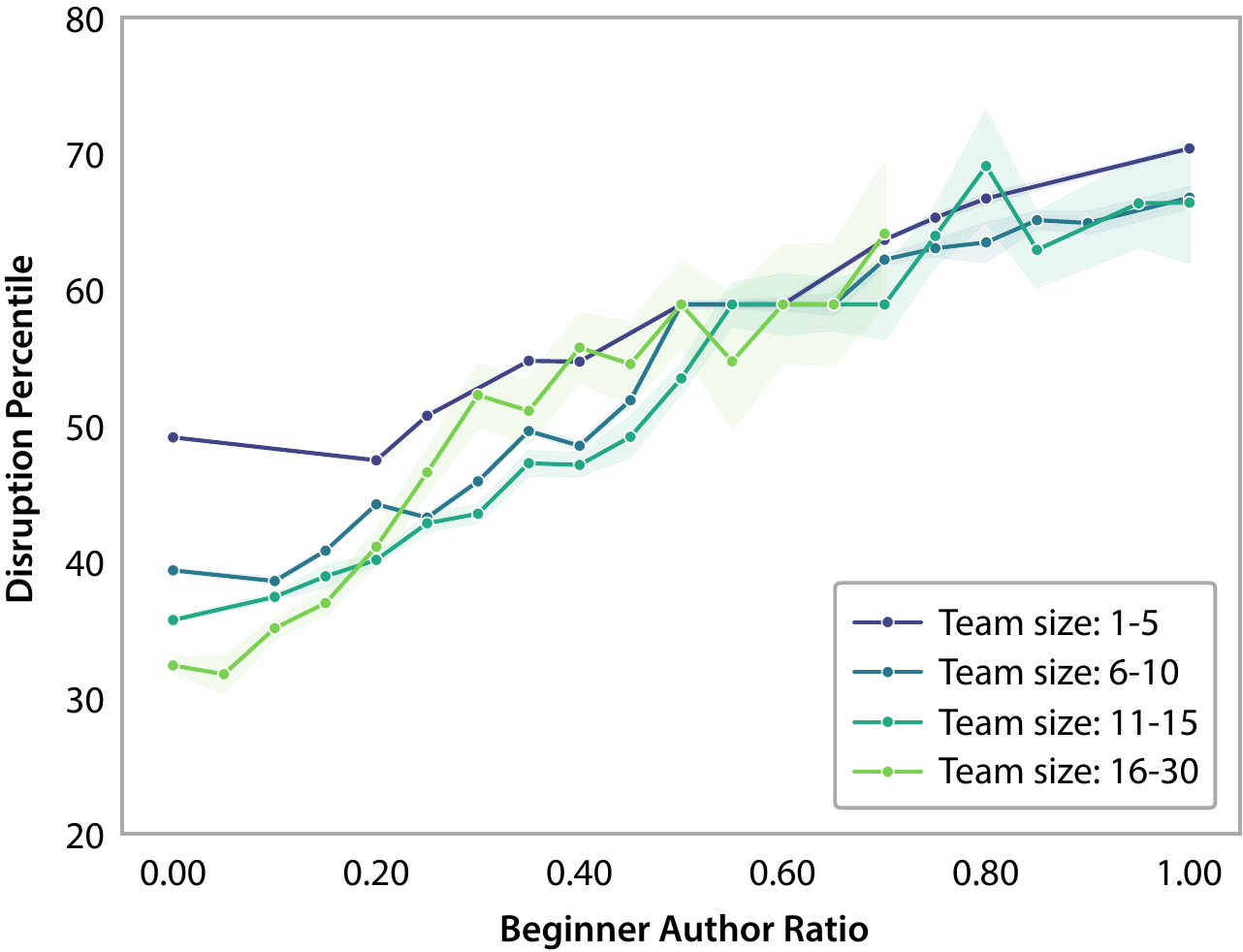}
\caption{\textbf{Positive correlation between the proportion of beginner authors and disruption percentile across varying team sizes.} As the beginner ratio increases, disruption levels rise across team sizes---and especially steeply so in large teams, where the usual size penalty attenuates. In fact, beginner-heavy large teams approach the disruption levels of beginner-heavy small teams.}\label{SIfig_s1}
\end{figure}

\clearpage

\begin{figure}
\centering
\hspace*{-0.14\linewidth}
\includegraphics[width=1.3\linewidth]{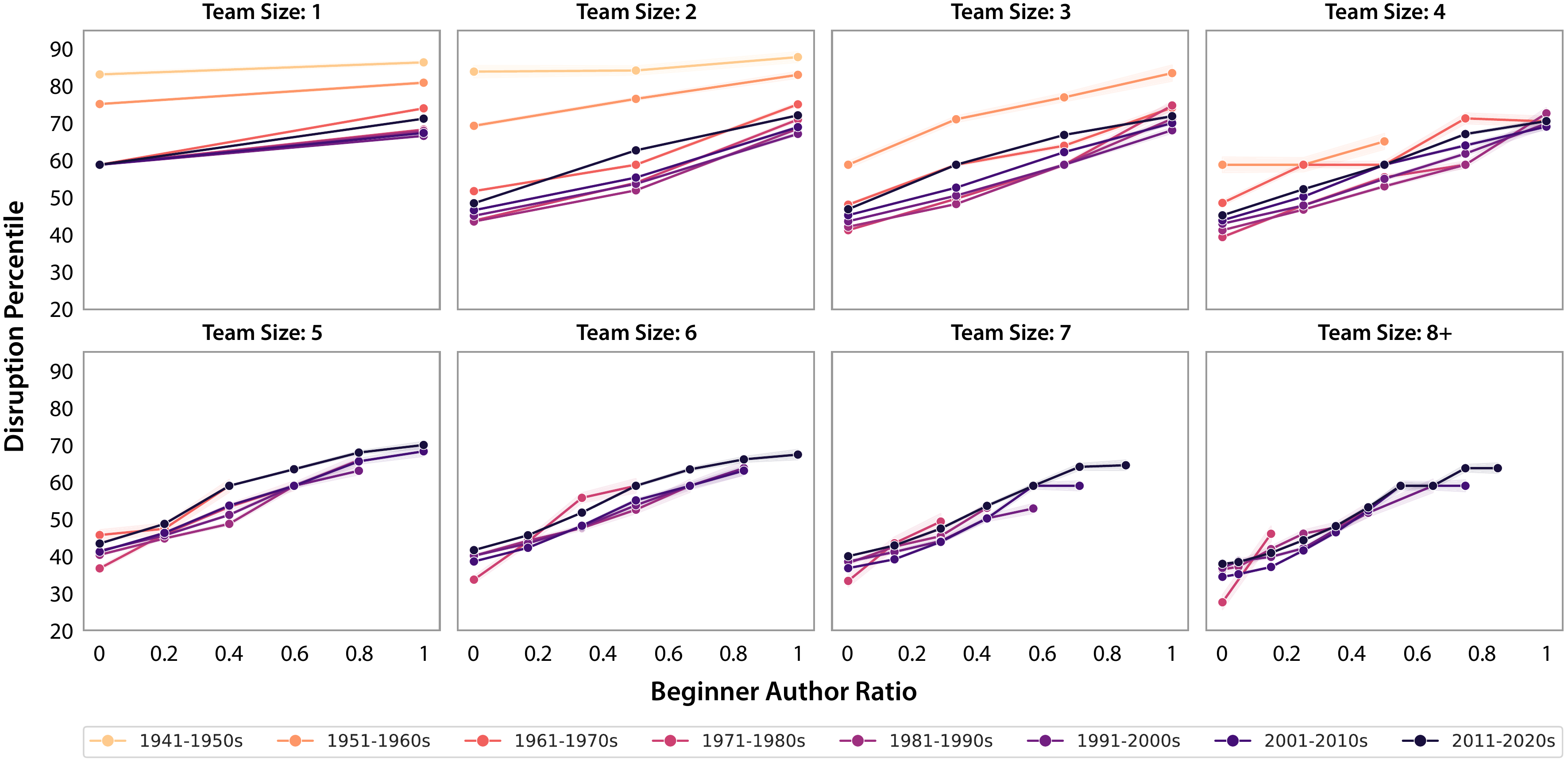}
\caption{\textbf{Beginner author ratio positively correlates with higher disruption percentiles across all decades.} This figure shows the relationship between the beginner author ratio and disruption percentile, broken down by team size and publication decade (1941–2020). We plot scatter points with at least $1000$ papers. Across all decades and team sizes, a higher share of beginner authors is associated with more disruptive work.}
\label{SIfig_s2}
\end{figure}

\clearpage

\begin{figure}
\centering
\hspace*{-0.18\linewidth}
\includegraphics[width=1.3\linewidth]{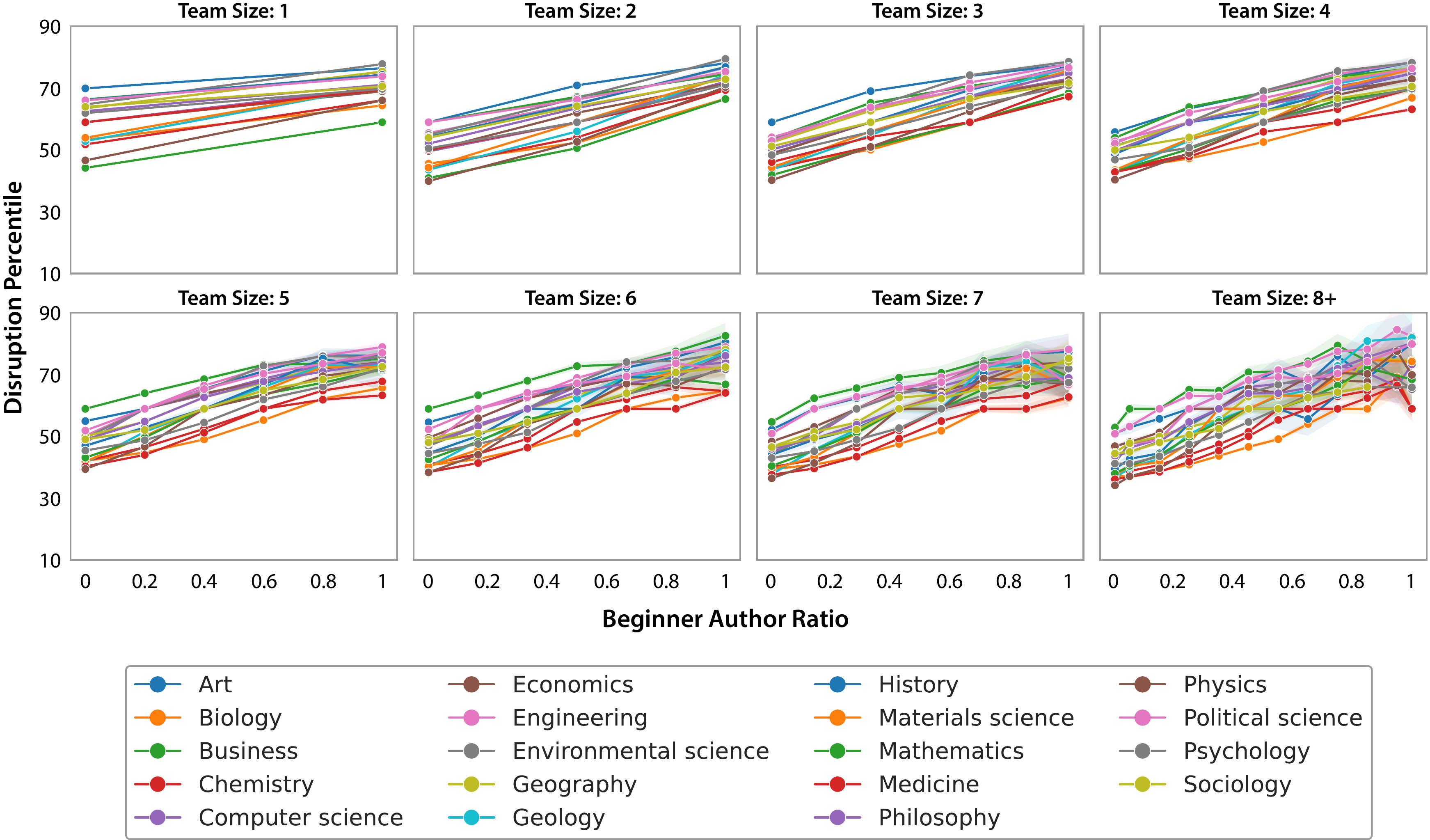}
\caption{\textbf{Beginner author ratio positively correlates with higher disruption percentiles across all 19 broad disciplines.} This figure shows the relationship between the beginner author ratio and disruption percentile, broken down by discipline and team size. Across disciplines, increasing the share of beginner authors correlates with higher disruption.}
\label{SIfig_s3}
\end{figure}

\clearpage

\begin{figure}
\centering
\hspace*{-0.14\linewidth}
\includegraphics[width=1.3\linewidth]{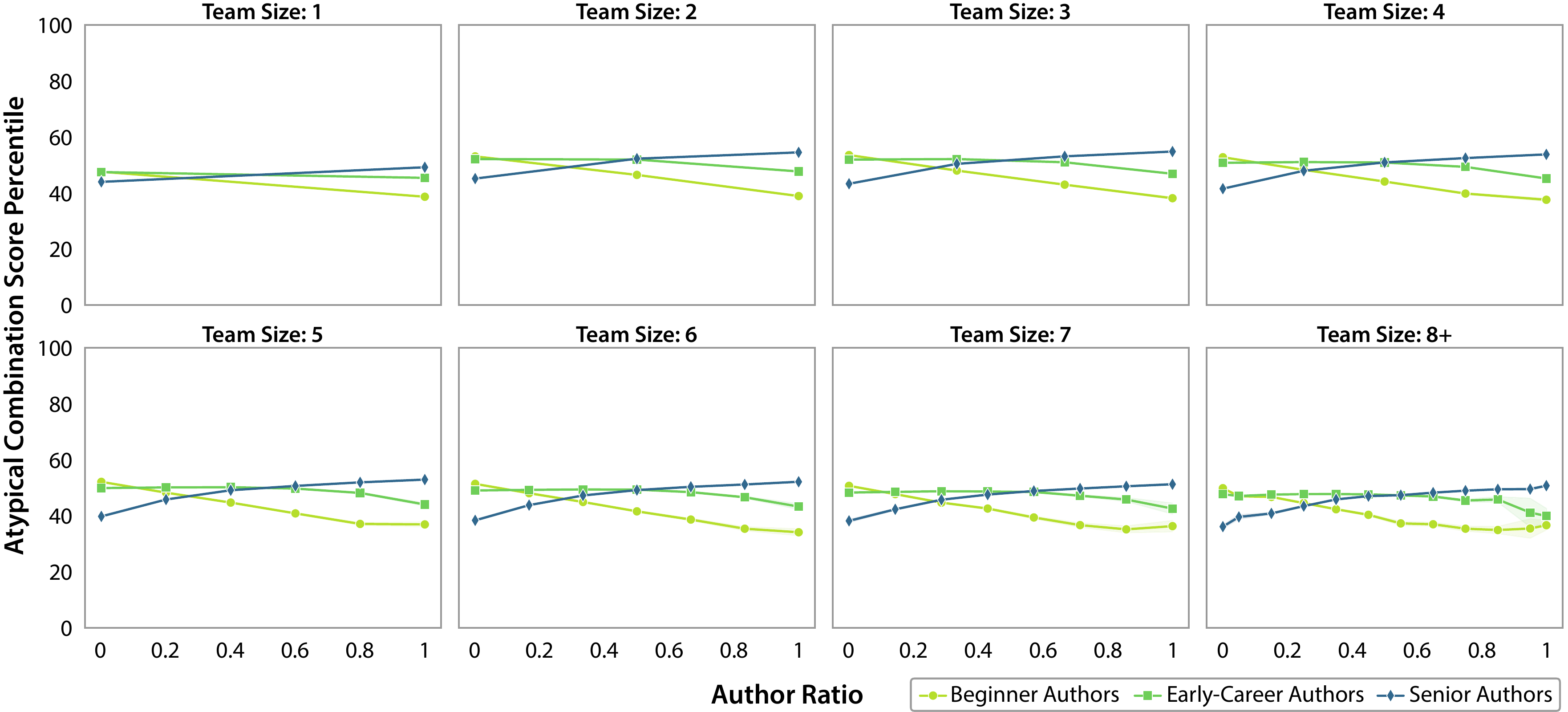}
\caption{\textbf{Beginner author ratio negatively correlates with atypical combination score percentiles (i.e., positively correlates with innovation) across team sizes.}
This figure demonstrates the negative correlation between the proportion of beginner authors and atypical combination score percentile across varying team sizes.}
\label{SIfig_s4}
\end{figure}

\clearpage

\begin{figure}
\centering
\hspace*{-0.18\linewidth}
\includegraphics[width=1.3\linewidth]{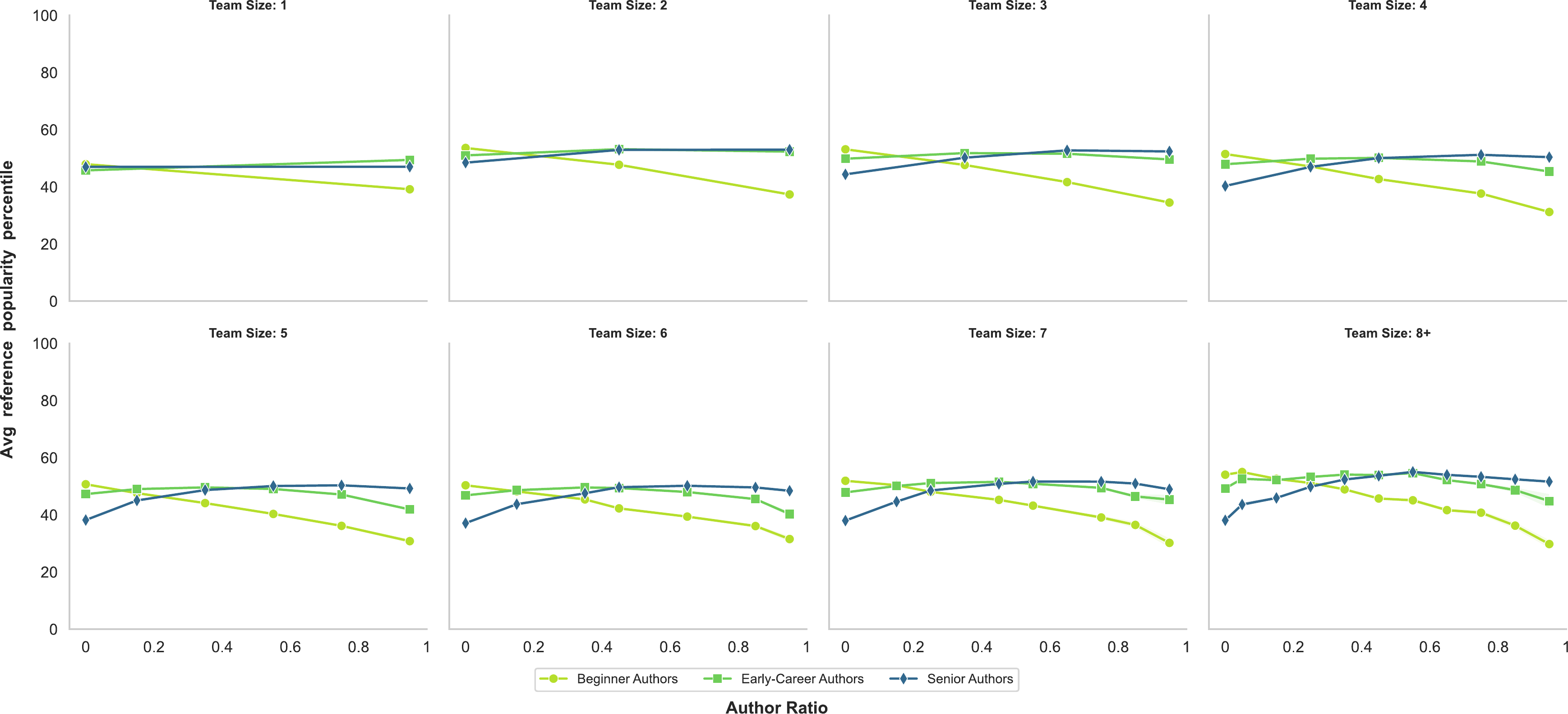}
\caption{\textbf{Beginner author ratio negatively correlates with average reference popularity across team sizes.}
This figure displays the ratio of beginner authors and their average reference popularity percentiles. Teams with a higher beginner ratio tend to cite less popular papers.}
\label{SIfig_s5}
\end{figure}

\clearpage

\begin{figure}
\centering
\includegraphics[width=0.9\textwidth]{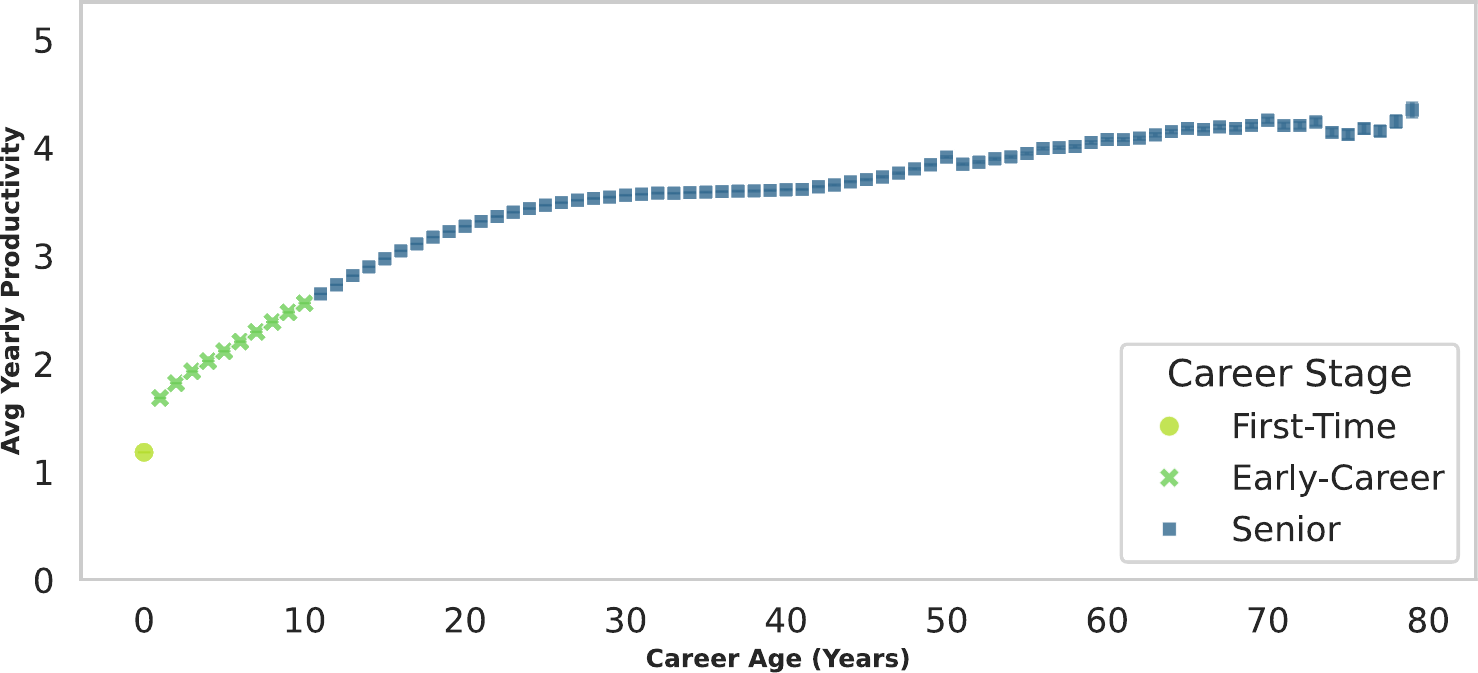}
\caption{\textbf{Positive relationship between career age and average yearly productivity.}
On average, beginner authors publish 1.18 papers in their debut year.
Their productivity rises markedly in the following year and then continues to increase gradually thereafter.}
\label{SIfig_s6}
\end{figure}

\clearpage

\begin{figure}
\centering
\hspace*{-0.18\linewidth}
\includegraphics[width=1.3\linewidth]{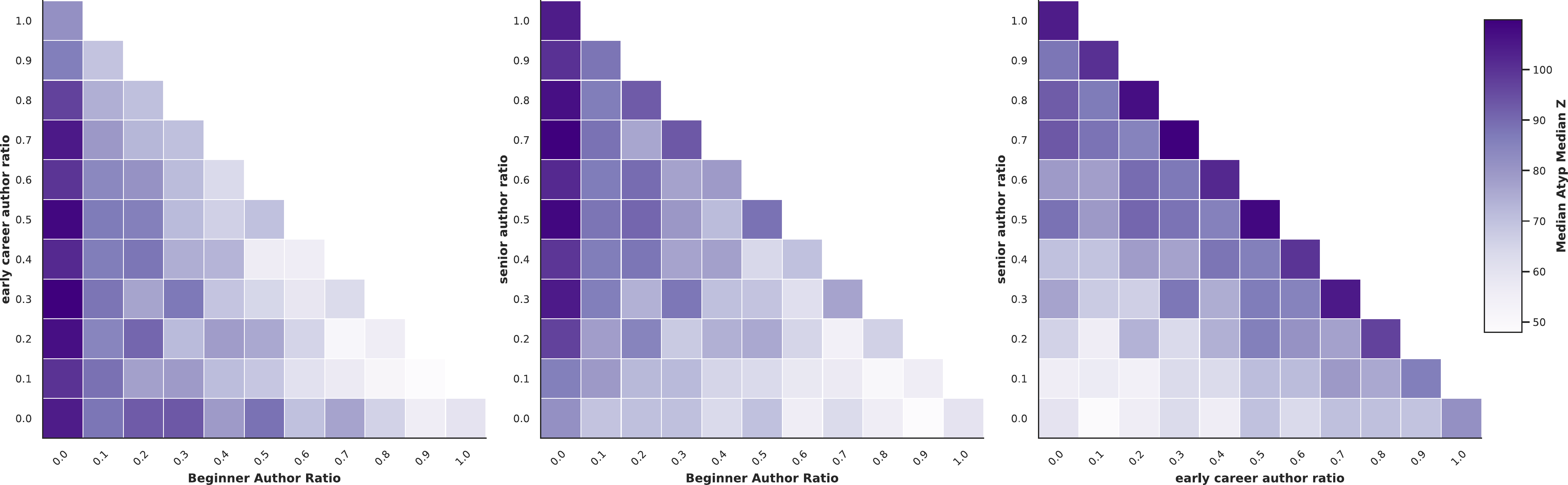}
\caption{\textbf{Early-career collaborators are associated with higher innovation in beginner-heavy teams.} We illustrate atypical combination scores of pairwise combinations of beginner, early-career, and senior authors across different author ratios. The heatmap cells visualize median atypical combination score percentiles, with lighter shades indicating higher innovation. While teams heavier on beginners are progressively more innovative, early-career collaborators are associated with better beginner innovation than senior collaborators.}
\label{SIfig_s7}
\end{figure}

\clearpage

\begin{figure}
\centering
\hspace*{-0.14\linewidth}
\includegraphics[width=1.3\linewidth]{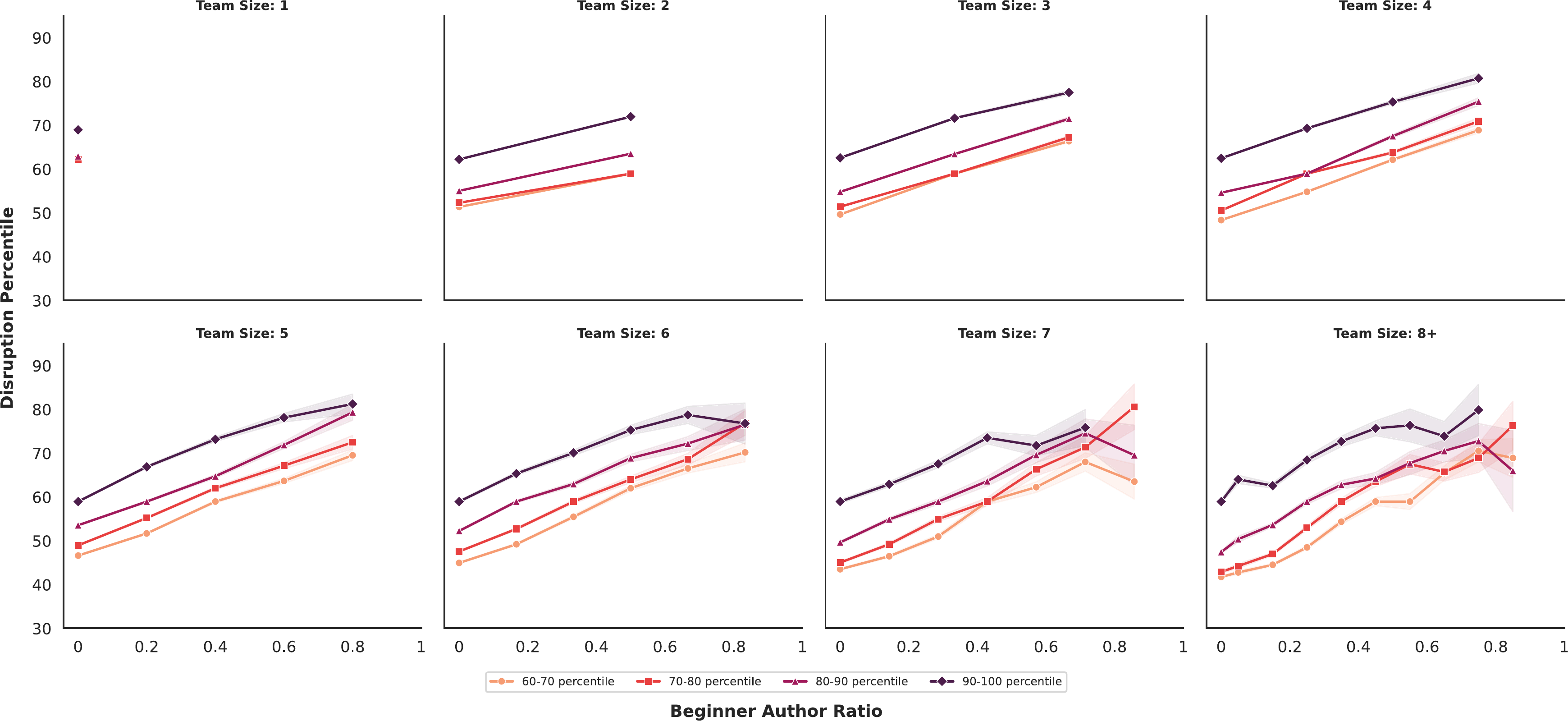}
\caption{\textbf{Impact of beginner author ratio on disruption across co-authors' disruption-ranked teams of varying sizes.}
This figure illustrates how disruption percentiles change with varying proportions of beginner authors in research teams, stratified by team size and collaborators' mean disruption percentiles.
Teams with an equal number of beginner authors become more disruptive as the disruption ranks of their collaborators increase.}
\label{SIfig_s8}
\end{figure}

\clearpage

\begin{figure}
\centering
\hspace*{-0.18\linewidth}
\includegraphics[width=1.3\linewidth]{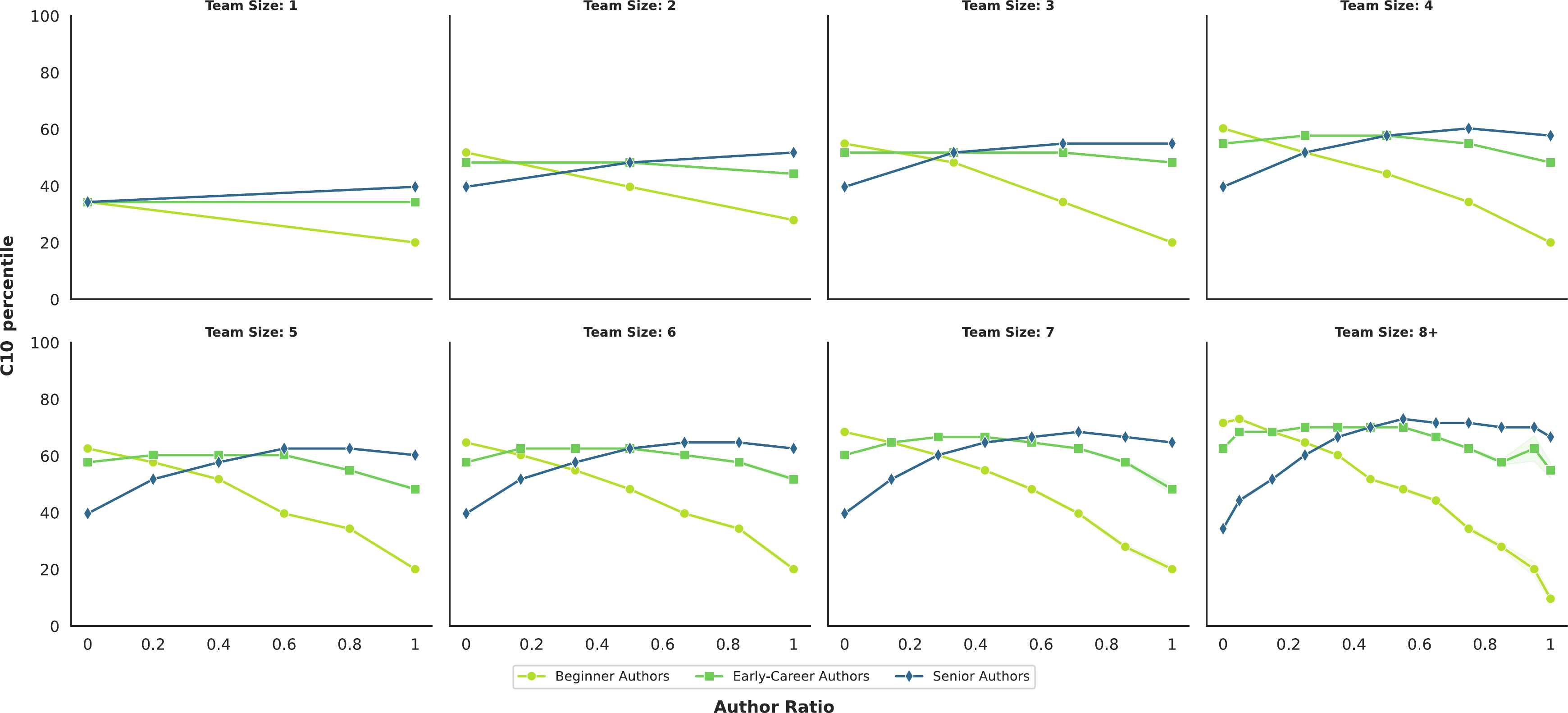}
\caption{\textbf{Beginner author ratio is associated with a negative citation impact in the overall dataset.} This figure shows the relationship between author career stage ratio and $C_{10}$ percentile (citations after 10 years) across team sizes. Increasing the proportion of beginner and early-career authors is generally associated with lower long-term citation impact.}
\label{SIfig_s9}
\end{figure}

\clearpage

\begin{figure}
\centering
\includegraphics[width=0.7\textwidth]{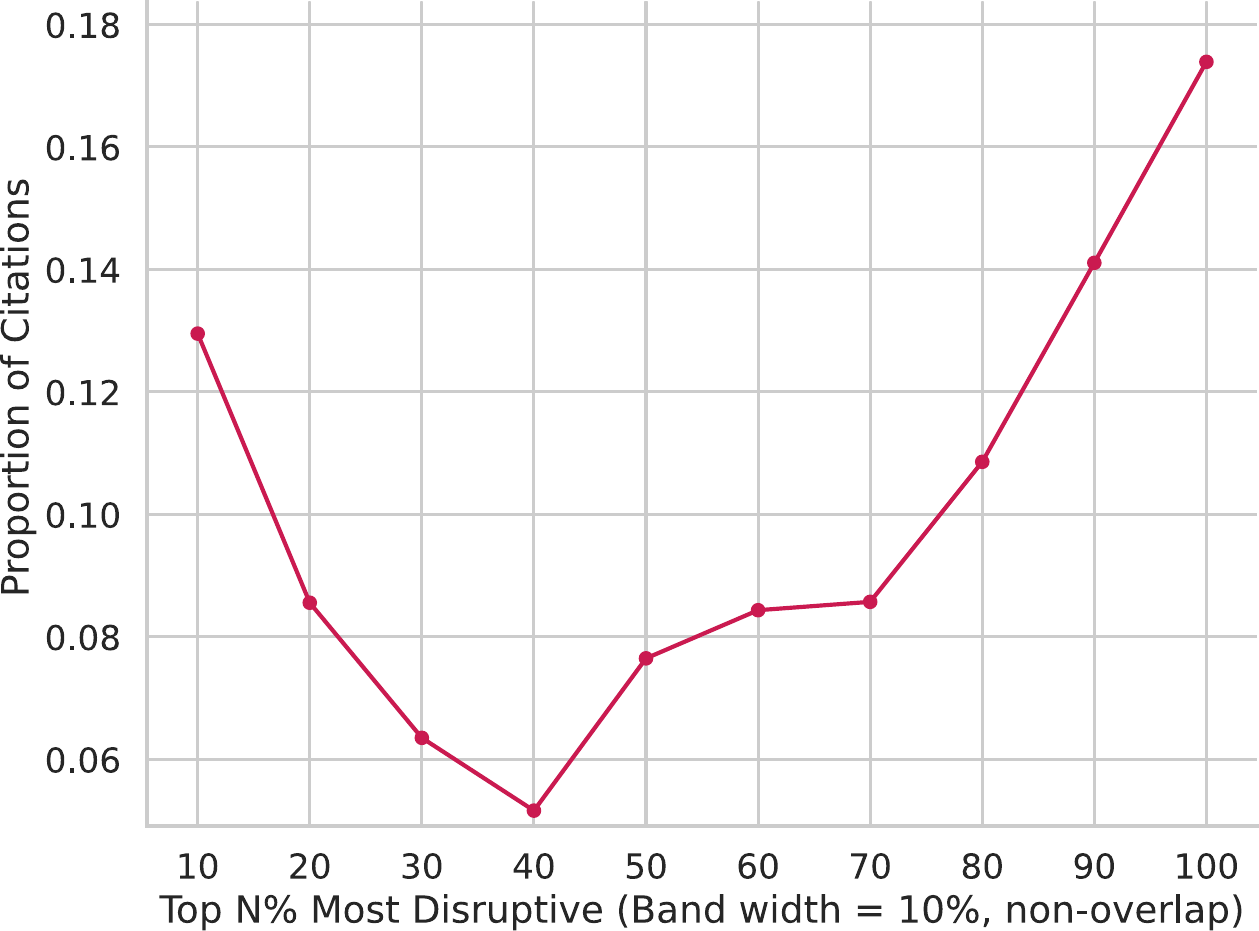}
\caption{Proportions of citations received by each of the top-$N$\% (non-cumulative) bin of the disruptive-ranked papers.}
\label{SIfig_s10}
\end{figure}

\clearpage

\begin{figure}
\centering
\hspace*{-0.18\linewidth}
\includegraphics[width=1.3\linewidth]{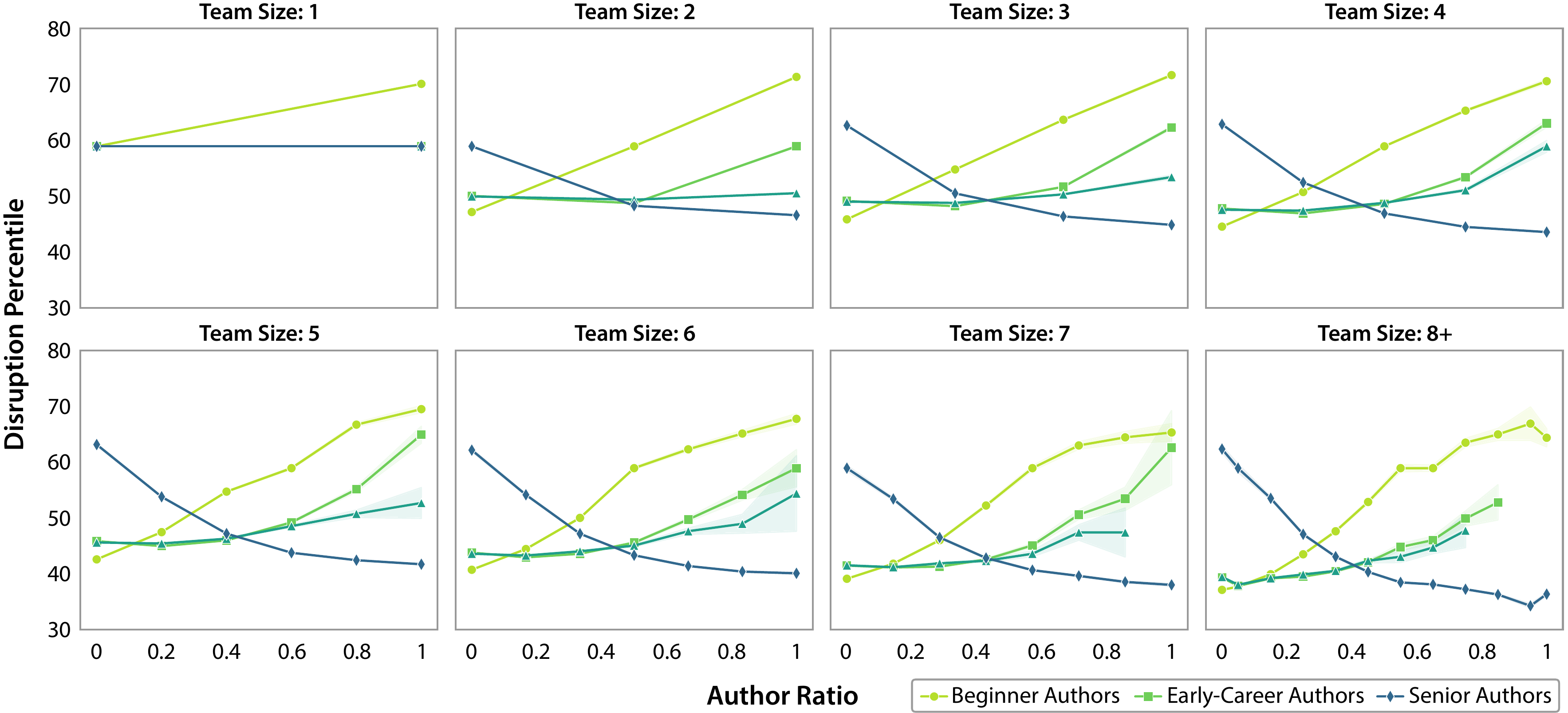}
\caption{\textbf{Beginner-heavy teams are disruptive across team sizes, robustly for a 4-way career stage split.} We show the relationship between disruption percentiles and author ratios at different career stages, split by team sizes. Beginner author ratios are positively associated with higher levels of disruption, while teams with higher senior author ratios tend to produce less disruptive work. The results are robust to team size. Shaded regions denote $95$\% C.I.}
\label{SIfig_s11}
\end{figure}

\clearpage

\begin{figure}
\centering
\hspace*{-0.14\linewidth}
\includegraphics[width=1.3\linewidth]{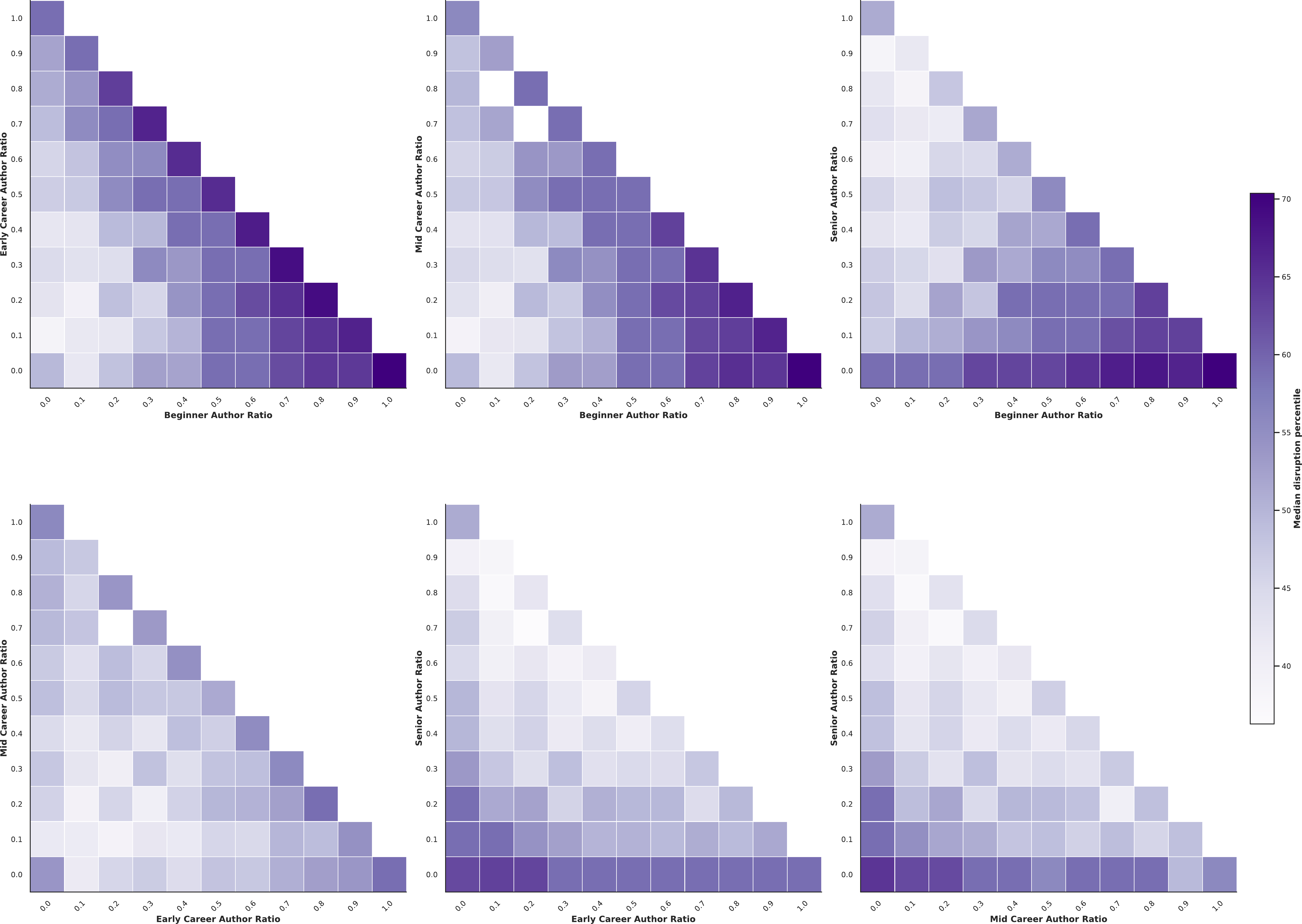}
\caption{\textbf{Early-career collaborators are associated with higher disruption in beginner-heavy teams under a four-way career-age split.} We illustrate disruptions of pairwise combinations of beginner, early-career, mid-career, and senior authors across different author ratios. The heatmap cells visualize median disruption percentiles, with darker shades indicating higher disruption. While teams heavier on beginners are progressively more disruptive, early-career collaborators are associated with better beginner disruption than mid-career or senior collaborators. }
\label{SIfig_s12}
\end{figure}

\clearpage

\begin{figure}
\centering
\hspace*{-0.18\linewidth}
\includegraphics[width=1.3\linewidth]{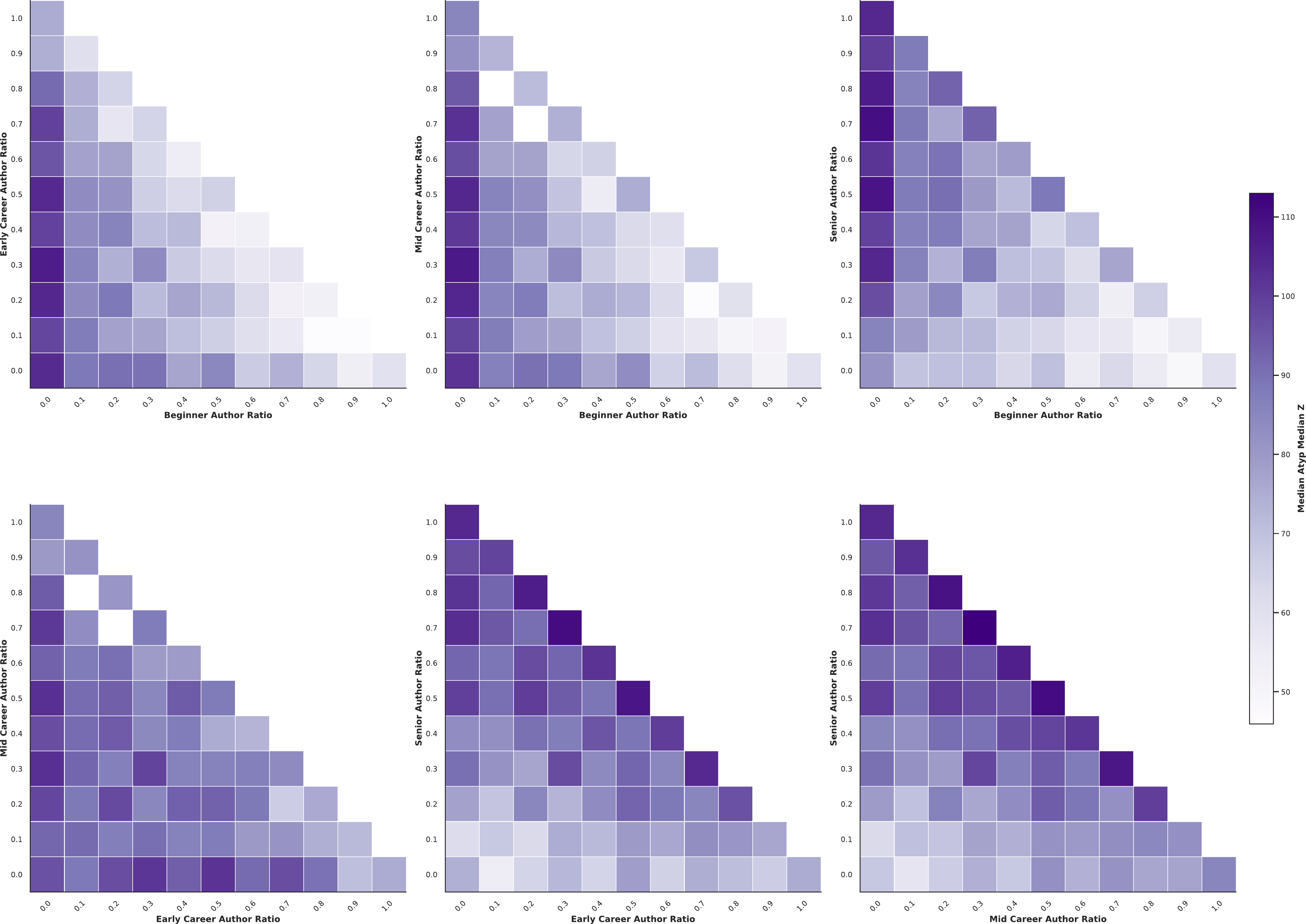}
\caption{\textbf{Early-career collaborators are associated with higher innovation in beginner-heavy teams under a four-way career-age split.} We illustrate atypical combination scores of pairwise combinations of beginner, early-career, mid-career, and senior authors across different author ratios. The heatmap cells visualize median atypical combination score percentiles, with lighter shades indicating higher innovation. While teams heavier on beginners are progressively more innovative, early-career collaborators are associated with better beginner innovation than mid-career or senior collaborators.}
\label{SIfig_s13}
\end{figure}

\clearpage

\begin{table}\centering
\caption{\textbf{Kendall’s $\tau$ between author ratio and disruption percentile by team size and career stage.} Entries are Kendall’s $\tau$ with bootstrap 95\% percentile CIs in brackets; significance stars reflect adjusted $P$-values: $^{*}P<0.05$, $^{**}P<0.01$, $^{***}P<0.001$. Benjamini–Hochberg FDR correction applied across all tests.}
\begin{tabular}{llll}
Team size & Beginner & Early-career & Senior \\
\midrule
1 & 1.000 [1.000, 1.000] & -- & -- \\
2 & 1.000 [1.000, 1.000] & 0.333 [-1.000, 1.000] & -1.000 [-1.000, -1.000] \\
3 & 1.000 [1.000, 1.000] & 0.667 [-1.000, 1.000] & -1.000 [-1.000, -1.000] \\
4 & 1.000 [1.000, 1.000]* & 0.600 [-0.500, 1.000] & -1.000 [-1.000, -1.000]* \\
5 & 1.000 [1.000, 1.000]** & 0.733 [-0.091, 1.000] & -1.000 [-1.000, -1.000]** \\
6 & 1.000 [1.000, 1.000]** & 0.619 [-0.200, 1.000] & -1.000 [-1.000, -1.000]** \\
7 & 1.000 [1.000, 1.000]*** & 0.786 [0.200, 1.000]* & -1.000 [-1.000, -1.000]*** \\
8+ & 0.931 [0.738, 1.000]*** & 0.848 [0.517, 1.000]*** & -0.939 [-1.000, -0.714]*** \\
\bottomrule
\end{tabular}
\label{SItab_s1}
\end{table}

\clearpage

\begin{table}
\centering
\caption{\textbf{Correlations between beginner author ratio and disruption percentile across decades in the overall dataset.} All reported correlations are statistically significant at $P < 0.001$ (denoted by ***), demonstrating the robustness of this association over time.}
\begin{tabular}{lrrrr}
Group & $N$ & Pearson $r$ & Spearman $\rho$ & Kendall $\tau$ \\
\midrule
All Data & 29,054,261 & 0.107232*** & 0.081831*** & 0.062813*** \\
1941-1950 & 47,213 & 0.042378*** & 0.048536*** & 0.038080*** \\
1951-1960 & 235,753 & 0.046042*** & 0.043221*** & 0.033310*** \\
1961-1970 & 730,667 & 0.072997*** & 0.058709*** & 0.045113*** \\
1971-1980 & 1,477,240 & 0.068561*** & 0.053265*** & 0.041002*** \\
1981-1990 & 2,418,102 & 0.073413*** & 0.052354*** & 0.040219*** \\
1991-2000 & 3,757,344 & 0.075212*** & 0.051006*** & 0.038985*** \\
2001-2010 & 7,290,170 & 0.096032*** & 0.067303*** & 0.051502*** \\
2011-2020 & 13,097,772 & 0.137403*** & 0.109909*** & 0.084223*** \\
\bottomrule
\end{tabular}
\label{SItab_s2}
\end{table}

\clearpage

\begin{table*}[t]\centering
\caption{\textbf{Correlations between author beginner ratio and disruption percentile by discipline in the overall dataset.} Pearson’s $r$, Spearman’s $\rho$, and Kendall’s $\tau$ with significance stars: $^{***}P<0.001$.}
\scriptsize
\setlength{\tabcolsep}{5pt}
\begin{tabular}{lrrrr}
Group & $N$ & Pearson $r$ & Spearman $\rho$ & Kendall $\tau$ \\
\midrule
Art & 1{,}106{,}563 & 0.0850$^{***}$ & 0.0751$^{***}$ & 0.0585$^{***}$ \\
Biology & 8{,}240{,}399 & 0.0972$^{***}$ & 0.0679$^{***}$ & 0.0519$^{***}$ \\
Business & 1{,}721{,}427 & 0.1263$^{***}$ & 0.1255$^{***}$ & 0.0974$^{***}$ \\
Chemistry & 7{,}740{,}170 & 0.0976$^{***}$ & 0.0797$^{***}$ & 0.0611$^{***}$ \\
Computer science & 9{,}102{,}699 & 0.1193$^{***}$ & 0.1118$^{***}$ & 0.0865$^{***}$ \\
Economics & 2{,}670{,}974 & 0.1196$^{***}$ & 0.1115$^{***}$ & 0.0870$^{***}$ \\
Engineering & 5{,}735{,}626 & 0.1162$^{***}$ & 0.1048$^{***}$ & 0.0804$^{***}$ \\
Environmental science & 1{,}418{,}544 & 0.1424$^{***}$ & 0.1297$^{***}$ & 0.0995$^{***}$ \\
Geography & 1{,}710{,}299 & 0.1248$^{***}$ & 0.1115$^{***}$ & 0.0862$^{***}$ \\
Geology & 1{,}875{,}846 & 0.1351$^{***}$ & 0.1218$^{***}$ & 0.0943$^{***}$ \\
History & 713{,}100 & 0.0751$^{***}$ & 0.0573$^{***}$ & 0.0450$^{***}$ \\
Materials science & 5{,}159{,}780 & 0.1025$^{***}$ & 0.0883$^{***}$ & 0.0680$^{***}$ \\
Mathematics & 5{,}968{,}273 & 0.1184$^{***}$ & 0.1140$^{***}$ & 0.0891$^{***}$ \\
Medicine & 7{,}501{,}956 & 0.0867$^{***}$ & 0.0558$^{***}$ & 0.0421$^{***}$ \\
Philosophy & 3{,}129{,}726 & 0.1066$^{***}$ & 0.0943$^{***}$ & 0.0736$^{***}$ \\
Physics & 8{,}476{,}490 & 0.1141$^{***}$ & 0.0999$^{***}$ & 0.0774$^{***}$ \\
Political science & 2{,}898{,}145 & 0.1007$^{***}$ & 0.0917$^{***}$ & 0.0720$^{***}$ \\
Psychology & 3{,}472{,}859 & 0.1119$^{***}$ & 0.0872$^{***}$ & 0.0672$^{***}$ \\
Sociology & 2{,}462{,}381 & 0.0983$^{***}$ & 0.0747$^{***}$ & 0.0582$^{***}$ \\
\bottomrule
\end{tabular}
\label{SItab_s3}
\end{table*}

\clearpage

\begin{table}\centering
\caption{\textbf{Kendall’s $\tau$ between author ratio and mean reference popularity percentile by team size and career stage.} Entries are Kendall’s $\tau$ with bootstrap 95\% percentile CIs in brackets; significance stars reflect adjusted $P$-values: $^{*}P<0.05$, $^{**}P<0.01$, $^{***}P<0.001$. Benjamini–Hochberg FDR correction applied across all tests.}
\begin{tabular}{llll}
Team size & Beginner & Early-career & Senior \\
\midrule
1 & -1.000 [-1.000, -1.000] & 1.000 [1.000, 1.000] & 1.000 [1.000, 1.000] \\
2 & -1.000 [-1.000, -1.000] & -0.333 [-1.000, 1.000] & 1.000 [1.000, 1.000] \\
3 & -1.000 [-1.000, -1.000] & -0.333 [-1.000, 1.000] & 0.667 [-1.000, 1.000] \\
4 & -1.000 [-1.000, -1.000] & -0.200 [-1.000, 1.000] & 0.800 [0.000, 1.000] \\
5 & -1.000 [-1.000, -1.000]* & -0.467 [-1.000, 0.538] & 0.733 [-0.091, 1.000] \\
6 & -1.000 [-1.000, -1.000]** & -0.429 [-1.000, 0.468] & 0.524 [-0.333, 1.000] \\
7 & -1.000 [-1.000, -1.000]*** & -0.357 [-1.000, 0.500] & 0.500 [-0.391, 1.000] \\
8+ & -0.970 [-1.000, -0.811]*** & -0.333 [-0.806, 0.310] & 0.485 [-0.153, 0.931] \\
\bottomrule
\end{tabular}
\label{SItab_s4}
\end{table}

\clearpage

\begin{sidewaystable}\centering
\caption{Disruption percentiles by author-ratio bins. Cells show median disruption percentile and sample size ($n$) for the left subplot of Figure 3: Early Career Author Ratio (rows) vs Beginner Author Ratio (columns).}
\scriptsize
\setlength\tabcolsep{4pt}
\renewcommand{\arraystretch}{1.1}
\begin{tabular}{lccccccccccc}
\toprule
\makecell{Early Career \\ Ratio} & \multicolumn{11}{c}{Beginner Author Ratio (bin center)} \\
\cmidrule(lr){2-12}
& 0.0 & 0.1 & 0.2 & 0.3 & 0.4 & 0.5 & 0.6 & 0.7 & 0.8 & 0.9 & 1.0 \\
\midrule
1.0 & \makecell{58.92 \\ \scriptsize n=3.34M} & -- & -- & -- & -- & -- & -- & -- & -- & -- & -- \\
0.9 & \makecell{47.07 \\ \scriptsize n=33.60K} & \makecell{58.92 \\ \scriptsize n=9.00K} & -- & -- & -- & -- & -- & -- & -- & -- & -- \\
0.8 & \makecell{47.79 \\ \scriptsize n=396.08K} & \makecell{48.60 \\ \scriptsize n=3.85K} & \makecell{58.92 \\ \scriptsize n=96.58K} & -- & -- & -- & -- & -- & -- & -- & -- \\
0.7 & \makecell{46.55 \\ \scriptsize n=1.12M} & \makecell{49.06 \\ \scriptsize n=43.26K} & \makecell{50.63 \\ \scriptsize n=1.52K} & \makecell{62.87 \\ \scriptsize n=251.87K} & -- & -- & -- & -- & -- & -- & -- \\
0.6 & \makecell{42.81 \\ \scriptsize n=340.80K} & \makecell{43.87 \\ \scriptsize n=43.83K} & \makecell{51.37 \\ \scriptsize n=99.33K} & \makecell{51.86 \\ \scriptsize n=9.75K} & \makecell{62.96 \\ \scriptsize n=25.23K} & -- & -- & -- & -- & -- & -- \\
0.5 & \makecell{45.49 \\ \scriptsize n=3.69M} & \makecell{44.80 \\ \scriptsize n=103.52K} & \makecell{52.44 \\ \scriptsize n=286.04K} & \makecell{54.32 \\ \scriptsize n=28.60K} & \makecell{54.95 \\ \scriptsize n=3.28K} & \makecell{62.80 \\ \scriptsize n=657.12K} & -- & -- & -- & -- & -- \\
0.4 & \makecell{40.61 \\ \scriptsize n=633.04K} & \makecell{40.89 \\ \scriptsize n=105.69K} & \makecell{47.06 \\ \scriptsize n=222.45K} & \makecell{47.35 \\ \scriptsize n=24.77K} & \makecell{58.92 \\ \scriptsize n=82.69K} & \makecell{58.92 \\ \scriptsize n=2.85K} & \makecell{65.03 \\ \scriptsize n=22.20K} & -- & -- & -- & -- \\
0.3 & \makecell{43.51 \\ \scriptsize n=2.15M} & \makecell{41.98 \\ \scriptsize n=227.08K} & \makecell{42.05 \\ \scriptsize n=16.98K} & \makecell{53.26 \\ \scriptsize n=770.09K} & \makecell{52.00 \\ \scriptsize n=16.27K} & \makecell{58.92 \\ \scriptsize n=20.31K} & \makecell{58.92 \\ \scriptsize n=6.43K} & \makecell{66.95 \\ \scriptsize n=157.57K} & -- & -- & -- \\
0.2 & \makecell{42.06 \\ \scriptsize n=1.45M} & \makecell{39.05 \\ \scriptsize n=72.95K} & \makecell{47.61 \\ \scriptsize n=693.13K} & \makecell{44.43 \\ \scriptsize n=8.98K} & \makecell{52.38 \\ \scriptsize n=108.40K} & \makecell{58.92 \\ \scriptsize n=154.84K} & \makecell{58.92 \\ \scriptsize n=41.41K} & \makecell{58.92 \\ \scriptsize n=1.30K} & \makecell{67.96 \\ \scriptsize n=57.95K} & -- & -- \\
0.1 & \makecell{38.26 \\ \scriptsize n=424.27K} & \makecell{41.31 \\ \scriptsize n=230.57K} & \makecell{41.52 \\ \scriptsize n=26.76K} & \makecell{46.70 \\ \scriptsize n=87.66K} & \makecell{48.16 \\ \scriptsize n=23.10K} & \makecell{55.25 \\ \scriptsize n=28.70K} & \makecell{58.92 \\ \scriptsize n=10.24K} & \makecell{58.92 \\ \scriptsize n=16.39K} & \makecell{63.68 \\ \scriptsize n=2.26K} & \makecell{66.53 \\ \scriptsize n=8.00K} & -- \\
0.0 & \makecell{51.26 \\ \scriptsize n=6.90M} & \makecell{41.54 \\ \scriptsize n=102.26K} & \makecell{47.66 \\ \scriptsize n=389.98K} & \makecell{51.67 \\ \scriptsize n=598.23K} & \makecell{51.13 \\ \scriptsize n=65.92K} & \makecell{55.51 \\ \scriptsize n=1.11M} & \makecell{58.92 \\ \scriptsize n=31.56K} & \makecell{58.92 \\ \scriptsize n=202.78K} & \makecell{63.50 \\ \scriptsize n=62.56K} & \makecell{63.45 \\ \scriptsize n=7.95K} & \makecell{70.37 \\ \scriptsize n=1.10M} \\
\bottomrule
\end{tabular}
\label{SItab_s5}
\end{sidewaystable}

\clearpage

\begin{sidewaystable}\centering
\caption{Disruption percentiles by author-ratio bins. Cells show median disruption percentile and sample size ($n$) for the middle subplot of Figure 3: Senior Author Ratio (rows) vs Beginner Author Ratio (columns).}
\scriptsize
\setlength\tabcolsep{4pt}
\renewcommand{\arraystretch}{1.1}
\begin{tabular}{lccccccccccc}
\toprule
\makecell{Senior Author \\ Ratio} & \multicolumn{11}{c}{Beginner Author Ratio (bin center)} \\
\cmidrule(lr){2-12}
& 0.0 & 0.1 & 0.2 & 0.3 & 0.4 & 0.5 & 0.6 & 0.7 & 0.8 & 0.9 & 1.0 \\
\midrule
1.0 & \makecell{51.28 \\ \scriptsize n=6.89M} & -- & -- & -- & -- & -- & -- & -- & -- & -- & -- \\
0.9 & \makecell{38.28 \\ \scriptsize n=420.01K} & \makecell{41.68 \\ \scriptsize n=98.80K} & -- & -- & -- & -- & -- & -- & -- & -- & -- \\
0.8 & \makecell{42.07 \\ \scriptsize n=1.45M} & \makecell{38.52 \\ \scriptsize n=52.13K} & \makecell{47.69 \\ \scriptsize n=388.24K} & -- & -- & -- & -- & -- & -- & -- & -- \\
0.7 & \makecell{43.49 \\ \scriptsize n=2.16M} & \makecell{41.49 \\ \scriptsize n=212.58K} & \makecell{41.05 \\ \scriptsize n=12.39K} & \makecell{51.68 \\ \scriptsize n=597.27K} & -- & -- & -- & -- & -- & -- & -- \\
0.6 & \makecell{40.58 \\ \scriptsize n=636.34K} & \makecell{39.94 \\ \scriptsize n=138.23K} & \makecell{45.13 \\ \scriptsize n=260.45K} & \makecell{44.72 \\ \scriptsize n=32.75K} & \makecell{51.15 \\ \scriptsize n=65.34K} & -- & -- & -- & -- & -- & -- \\
0.5 & \makecell{45.48 \\ \scriptsize n=3.69M} & \makecell{42.67 \\ \scriptsize n=169.50K} & \makecell{48.74 \\ \scriptsize n=453.76K} & \makecell{47.59 \\ \scriptsize n=58.25K} & \makecell{45.61 \\ \scriptsize n=8.19K} & \makecell{55.51 \\ \scriptsize n=1.11M} & -- & -- & -- & -- & -- \\
0.4 & \makecell{42.75 \\ \scriptsize n=343.40K} & \makecell{41.40 \\ \scriptsize n=90.53K} & \makecell{46.92 \\ \scriptsize n=226.22K} & \makecell{45.30 \\ \scriptsize n=38.30K} & \makecell{52.09 \\ \scriptsize n=120.37K} & \makecell{51.59 \\ \scriptsize n=4.38K} & \makecell{58.92 \\ \scriptsize n=31.20K} & -- & -- & -- & -- \\
0.3 & \makecell{46.53 \\ \scriptsize n=1.12M} & \makecell{45.28 \\ \scriptsize n=114.33K} & \makecell{43.44 \\ \scriptsize n=10.34K} & \makecell{53.44 \\ \scriptsize n=759.59K} & \makecell{51.46 \\ \scriptsize n=16.98K} & \makecell{55.59 \\ \scriptsize n=25.59K} & \makecell{55.22 \\ \scriptsize n=8.00K} & \makecell{58.92 \\ \scriptsize n=202.46K} & -- & -- & -- \\
0.2 & \makecell{47.78 \\ \scriptsize n=396.88K} & \makecell{44.10 \\ \scriptsize n=13.56K} & \makecell{52.24 \\ \scriptsize n=379.84K} & \makecell{47.81 \\ \scriptsize n=4.52K} & \makecell{58.92 \\ \scriptsize n=77.20K} & \makecell{58.92 \\ \scriptsize n=154.64K} & \makecell{58.92 \\ \scriptsize n=41.77K} & \makecell{58.92 \\ \scriptsize n=1.44K} & \makecell{63.50 \\ \scriptsize n=62.31K} & -- & -- \\
0.1 & \makecell{47.02 \\ \scriptsize n=33.92K} & \makecell{49.62 \\ \scriptsize n=43.17K} & \makecell{50.89 \\ \scriptsize n=4.72K} & \makecell{53.86 \\ \scriptsize n=37.16K} & \makecell{55.55 \\ \scriptsize n=11.35K} & \makecell{58.92 \\ \scriptsize n=22.18K} & \makecell{58.92 \\ \scriptsize n=8.42K} & \makecell{61.97 \\ \scriptsize n=16.32K} & \makecell{63.19 \\ \scriptsize n=2.25K} & \makecell{63.38 \\ \scriptsize n=7.71K} & -- \\
0.0 & \makecell{58.92 \\ \scriptsize n=3.34M} & \makecell{58.92 \\ \scriptsize n=9.18K} & \makecell{58.92 \\ \scriptsize n=96.80K} & \makecell{62.87 \\ \scriptsize n=252.11K} & \makecell{62.93 \\ \scriptsize n=25.46K} & \makecell{62.80 \\ \scriptsize n=657.25K} & \makecell{64.96 \\ \scriptsize n=22.48K} & \makecell{66.93 \\ \scriptsize n=157.84K} & \makecell{67.95 \\ \scriptsize n=58.22K} & \makecell{66.47 \\ \scriptsize n=8.24K} & \makecell{70.37 \\ \scriptsize n=1.10M} \\
\bottomrule
\end{tabular}
\label{SItab_s6}
\end{sidewaystable}
            
\clearpage

\begin{sidewaystable}\centering
\caption{Disruption percentiles by author-ratio bins. Cells show median disruption percentile and sample size ($n$) for the right subplot of Figure 3: Senior Author Ratio (rows) vs Early Career Author Ratio (columns).}
\scriptsize
\setlength\tabcolsep{4pt}
\renewcommand{\arraystretch}{1.1}
\begin{tabular}{lccccccccccc}
\toprule
\makecell{Sen. Auth. \\ Ratio} & \multicolumn{11}{c}{Early Career Author Ratio (bin center)} \\
\cmidrule(lr){2-12}
& 0.0 & 0.1 & 0.2 & 0.3 & 0.4 & 0.5 & 0.6 & 0.7 & 0.8 & 0.9 & 1.0 \\
\midrule
1.0 & \makecell{51.28 \\ \scriptsize n=6.89M} & -- & -- & -- & -- & -- & -- & -- & -- & -- & -- \\
0.9 & \makecell{41.19 \\ \scriptsize n=105.89K} & \makecell{38.33 \\ \scriptsize n=412.91K} & -- & -- & -- & -- & -- & -- & -- & -- & -- \\
0.8 & \makecell{47.60 \\ \scriptsize n=391.71K} & \makecell{38.03 \\ \scriptsize n=60.02K} & \makecell{42.13 \\ \scriptsize n=1.44M} & -- & -- & -- & -- & -- & -- & -- & -- \\
0.7 & \makecell{51.64 \\ \scriptsize n=599.01K} & \makecell{42.11 \\ \scriptsize n=192.55K} & \makecell{37.49 \\ \scriptsize n=43.62K} & \makecell{43.55 \\ \scriptsize n=2.14M} & -- & -- & -- & -- & -- & -- & -- \\
0.6 & \makecell{51.07 \\ \scriptsize n=66.31K} & \makecell{43.75 \\ \scriptsize n=47.90K} & \makecell{44.39 \\ \scriptsize n=286.62K} & \makecell{39.67 \\ \scriptsize n=107.06K} & \makecell{40.64 \\ \scriptsize n=625.24K} & -- & -- & -- & -- & -- & -- \\
0.5 & \makecell{55.51 \\ \scriptsize n=1.11M} & \makecell{47.53 \\ \scriptsize n=63.00K} & \makecell{48.84 \\ \scriptsize n=450.49K} & \makecell{43.45 \\ \scriptsize n=135.26K} & \makecell{39.65 \\ \scriptsize n=45.38K} & \makecell{45.50 \\ \scriptsize n=3.69M} & -- & -- & -- & -- & -- \\
0.4 & \makecell{58.92 \\ \scriptsize n=31.69K} & \makecell{50.21 \\ \scriptsize n=19.70K} & \makecell{51.88 \\ \scriptsize n=111.81K} & \makecell{44.54 \\ \scriptsize n=44.19K} & \makecell{45.61 \\ \scriptsize n=281.95K} & \makecell{41.06 \\ \scriptsize n=27.20K} & \makecell{42.82 \\ \scriptsize n=337.88K} & -- & -- & -- & -- \\
0.3 & \makecell{58.92 \\ \scriptsize n=202.82K} & \makecell{55.64 \\ \scriptsize n=32.60K} & \makecell{48.68 \\ \scriptsize n=4.48K} & \makecell{53.60 \\ \scriptsize n=752.09K} & \makecell{46.17 \\ \scriptsize n=29.63K} & \makecell{45.95 \\ \scriptsize n=81.63K} & \makecell{43.51 \\ \scriptsize n=36.95K} & \makecell{46.55 \\ \scriptsize n=1.12M} & -- & -- & -- \\
0.2 & \makecell{63.48 \\ \scriptsize n=62.63K} & \makecell{58.92 \\ \scriptsize n=3.72K} & \makecell{58.92 \\ \scriptsize n=193.39K} & \makecell{51.40 \\ \scriptsize n=3.48K} & \makecell{55.75 \\ \scriptsize n=78.11K} & \makecell{52.53 \\ \scriptsize n=284.47K} & \makecell{50.61 \\ \scriptsize n=105.71K} & \makecell{43.98 \\ \scriptsize n=5.05K} & \makecell{47.79 \\ \scriptsize n=395.59K} & -- & -- \\
0.1 & \makecell{63.35 \\ \scriptsize n=7.96K} & \makecell{62.15 \\ \scriptsize n=17.28K} & \makecell{58.92 \\ \scriptsize n=3.29K} & \makecell{58.92 \\ \scriptsize n=25.90K} & \makecell{58.92 \\ \scriptsize n=10.92K} & \makecell{54.63 \\ \scriptsize n=30.52K} & \makecell{51.44 \\ \scriptsize n=12.93K} & \makecell{49.77 \\ \scriptsize n=40.80K} & \makecell{48.08 \\ \scriptsize n=4.16K} & \makecell{47.07 \\ \scriptsize n=33.43K} & -- \\
0.0 & \makecell{70.36 \\ \scriptsize n=1.10M} & \makecell{66.60 \\ \scriptsize n=8.26K} & \makecell{67.90 \\ \scriptsize n=58.22K} & \makecell{66.93 \\ \scriptsize n=157.84K} & \makecell{64.96 \\ \scriptsize n=22.46K} & \makecell{62.80 \\ \scriptsize n=657.23K} & \makecell{62.95 \\ \scriptsize n=25.47K} & \makecell{62.86 \\ \scriptsize n=252.09K} & \makecell{58.92 \\ \scriptsize n=96.76K} & \makecell{58.92 \\ \scriptsize n=9.17K} & \makecell{58.92 \\ \scriptsize n=3.34M} \\
\bottomrule
\end{tabular}
\label{SItab_s7}
\end{sidewaystable}

\clearpage

\begin{table}\centering
\caption{\textbf{Kendall’s $\tau$ between author ratio and disruption percentile by team size and career stage (for the 4-way career-stage split case).} Entries are Kendall’s $\tau$ with bootstrap 95\% percentile CIs in brackets; significance stars reflect adjusted $P$-values: $^{*}P<0.05$, $^{**}P<0.01$, $^{***}P<0.001$. Benjamini–Hochberg FDR correction applied across all tests.}
\begin{tabular}{lrrrr}
Team size & Beginner & Early-career & Mid-career & Senior \\
\midrule
1 & 1.000 [1.000, 1.000] & -- & -- & -- \\
2 & 1.000 [1.000, 1.000] & 0.333 [-1.000, 1.000] & 0.333 [-1.000, 1.000] & -1.000 [-1.000, -1.000] \\
3 & 1.000 [1.000, 1.000] & 0.667 [-1.000, 1.000] & 0.667 [-1.000, 1.000] & -1.000 [-1.000, -1.000] \\
4 & 1.000 [1.000, 1.000]* & 0.800 [0.000, 1.000] & 0.800 [0.000, 1.000] & -1.000 [-1.000, -1.000]* \\
5 & 1.000 [1.000, 1.000]** & 0.867 [0.385, 1.000]* & 0.867 [0.385, 1.000]* & -1.000 [-1.000, -1.000]** \\
6 & 1.000 [1.000, 1.000]** & 0.810 [0.250, 1.000]* & 0.905 [0.556, 1.000]** & -1.000 [-1.000, -1.000]** \\
7 & 1.000 [1.000, 1.000]*** & 0.857 [0.410, 1.000]** & 0.810 [0.294, 1.000]* & -1.000 [-1.000, -1.000]*** \\
8+ & 0.931 [0.705, 1.000]*** & 0.911 [0.600, 1.000]*** & 0.889 [0.484, 1.000]** & -0.939 [-1.000, -0.729]*** \\
\bottomrule
\end{tabular}
\label{SItab_s8}
\end{table}




\end{appendices}

\clearpage

\bibliography{sn-bibliography}

\end{document}